\documentclass[a4paper,fleqn,usenatbib,useAMS]{mnras}
\date{\today}
\usepackage{graphicx}
\usepackage{amsbsy,amsmath}

\DeclareMathAlphabet\mathbfcal{OMS}{cmsy}{b}{n}
\newcommand{\bvec}[1]{\boldsymbol{#1}}

\def\ps{pseudo-caustic}
\def\pss{pseudo-caustics}

\newcommand\be            {\begin{equation}}
\newcommand\ee            {\end{equation}}
\newcommand\ba            {\begin{aligned}}
\newcommand\ea            {\end{aligned}}
\newcommand{\bvr}{\boldsymbol{r}}
\usepackage{verbatim}
\usepackage{ae,aecompl}
\usepackage{txfonts}

\title{Gravitational Lensing by Ring-Like Structures}

\author[E. Lake and Z. Zheng]{Ethan Lake$^{1}$\thanks{Contact e-mail: \href{lake@physics.utah.edu}{lake@physics.utah.edu}} and  Zheng Zheng$^1$\thanks{Contact e-mail: \href{zhengzheng@astro.utah.edu}{zhengzheng@astro.utah.edu}}\\
$^{1}$Department of Physics and Astronomy, University of Utah,115 South 1400 East, Salt Lake City, UT 84112, USA}

\date{\today}
\pubyear{2016}

\begin{document}
\label{firstpage}
\pagerange{\pageref{firstpage}--\pageref{lastpage}}
\maketitle

\begin{abstract}
We study a class of gravitational lensing systems consisting of an inclined 
ring/belt, with and without an added point mass at the centre. We show that
a common feature of such systems are so-called ``pseudo-caustics'', across which the 
magnification of a point source changes discontinuously and yet remains finite.
Such a magnification change can be associated with either a change in
image multiplicity or a sudden change in the size of a lensed image. 
The existence of pseudo-caustics and the complex interplay between them and the formal 
caustics (which correspond to points of infinite magnification) can lead to interesting 
consequences, such as truncated or open caustics and a non-conservation of total image parity. The origin of the pseudo-caustics is found to be the non-differentiability of the 
solutions to the lens equation across the ring/belt boundaries, with the 
pseudo-caustics corresponding to ring/belt boundaries mapped into the source plane. We provide 
a few illustrative examples to understand the pseudo-caustic features, and in a separate
paper consider a specific astronomical application of our results to study microlensing by
extrasolar asteroid belts. 
\end{abstract}


\section{Introduction}

Gravitational lensing has developed from an interesting novelty predicted by 
general relativity \citep{Einstein36} to an essential tool in modern 
astrophysics. It has been applied with great success to (among other things)
detect extrasolar planets \citep[e.g.][]{Mao91,Bond04,Batista14},  estimate 
the masses of galaxy clusters \citep[e.g.][]{Tyson98,Bradav09,Hoekstra13}, and determine 
the large-scale distribution of matter in the Universe 
\citep[e.g.][]{Wittman00,Vikram15}. In addition, the theoretical study of 
gravitational lensing systems is of interest in mathematics, namely in
catastrophe theory and Morse theory \citep[e.g.][]{Erdl93,Petters95,Petters01}. 

One common goal of examining lensing systems from a theoretical standpoint 
is the characterization of the magnification of a source at a given position
from the lens mapping. In particular, there exist singularities in the 
mapping which form caustics (curves or points) in the source plane, where 
the magnification is formally 
infinite. Studying the behaviour of these caustics allows for the determination of certain properties of 
the lens mapping, and provides information about the number of images a 
source generates. Normally the number of images of a source changes when, 
and only when, the source crosses a caustic curve, with the total parity of the number of images
conserved throughout the source plane. However, some lens 
models have been found to violate this principle, namely singular isothermal
spheres and ellipsoids \citep[e.g.][]{Kovner87,Evans98,Rhie10} and variants thereof
\citep{Wang97,Shin08,Tessore15}. These 
systems possess features called ``pseudo-caustics'', where the number 
of images of a source can change without the source crossing a caustic.

In this paper, we examine lensing systems consisting of circular rings and belts
at various inclinations, with or without a point mass at the centre.
We find that these systems possess more general types of \pss, characterized
by a discontinuous but finite change in the source plane magnification. They are either 
associated with a change in image multiplicity 
or with a sudden change in the size of an image. These \pss\
produce complex and interesting
magnification patterns, despite the relative simplicity of the lensing systems in question. 
We aim to obtain a basic understanding of such lensing systems. Although we mainly focus on 
the limiting case of a sharp belt with a discontinuous mass distribution, 
we also analyse smooth mass distributions for a better understanding of the
origin of the pseudo-caustics and the effect of the smoothing on the 
pseudo-caustics. We show that the \pss\ arise from 
discontinuities in the lensing mass distribution.

Although our interest here is mainly theoretical and phenomenological, 
these kind of lensing systems do exist in astronomy, with a relevant example being
the asteroid belt around a star. In a related paper \citep{Lake16}, aided by the results presented 
here, we study the microlensing signatures of extrasolar asteroid belts
and explore the possibility of detection with future surveys.

This paper is organized as follows. In Section 2, we briefly recall key 
gravitational lens equations and present the solutions to the lensing systems
we consider. In Section 3, we examine the magnification patterns and 
pseudo-caustic structures for each of the systems in consideration and 
discuss the origin of pseudo-caustics. 
We summarize our results and
discuss implications in Section 4. Several more technical results are 
presented in a collection of appendices.

\section{Solutions to the Lens Equation}

In all the geometries we consider in this paper, a background source is lensed
by a mass lying between the source and the observer. We denote the angular
position in the image and source plane as $\bvec{r}_I = (x_I, y_I)$ and
$\bvec{r}_S = (x_S, y_S)$, respectively. We write all angular vectors in units of the 
Einstein ring radius,
\be \label{eq:er}
\theta_E = \sqrt{\frac{4GM}{c^2}\frac{D_{LS}}{D_LD_S}},
\ee
where $D_{L}$, $D_S$, and $D_{LS}$ are the (angular diameter) distances 
between the observer and the lens, the observer and the source, and the lens 
and the source, respectively. The mass $M$ is the total mass of the lens 
system we consider.

The general lens equation can be written as
\be
\bvec{r}_S=\bvec{r}_I-\bvec{\alpha}(\bvr_I).
\ee
The normalized deflection angle $\bvec{\alpha}$ is found by integrating the
contribution over the projected mass distribution of the lens,
\be 
\label{eq:integral}
\bvec{\alpha}({\bvec{r}_I) = \frac{1}{\pi}\int
\textrm{d}^2{\bvec{r}_I'}} \kappa(\bvec{r}_I') \frac{\bvec{r}_I - \bvec{r}_I'}{\left| \bvec{r}_I - \bvec{r}_I' \right|^2} ,
\ee
where $\kappa(\bvec{r}_I)$ is the dimensionless surface mass density of the lens,
\be
\kappa(\bvec{r}_I) = \frac{\Sigma(\bvec{r}_I)}{\Sigma_{\rm cr}}.
\ee
The critical surface density $\Sigma_{\rm cr}$ is defined as
\be
\Sigma_{\rm cr} \equiv \frac{c^2D_S}{4\pi G D_L D_{LS}}.
\ee
For a circularly symmetric lens, $\Sigma_{\rm cr}$ is equal to the mean surface density inside 
the Einstein ring radius. 
{ The surface density $\kappa$ is connected to the projected 
gravitational lensing potential $\psi$ (i.e. deflection potential) through 
Poisson's equation, } 
\be \label{eq:poisson} \nabla_{\bvr_I}^2 \psi(\bvr_I) = 2\kappa(\bvr_I). \ee
The magnification of the image at the image position $\bvec{r}_I$ can be computed as
\be \label{eq:detmag}
\mu(\bvr_I) = \frac{1}{\textrm{det}(A)},
\ee
where $\textrm{det}(A)$ is the determinant of the lensing Jacobian $A_{ij}$, 
a symmetric matrix 
formed from the derivatives of the lensing deflection $\bvec{\alpha} = \alpha_x \bvec{\hat x}_I + \alpha_y \bvec{\hat y}_I$,
\be
A_{ij} = \delta_{ij} - \frac{\partial \alpha_i}{\partial x_j},
\ee
with $x_1=x_I$ and $x_2=y_I$.
The total magnification for multiple images is obtained by $\mu_{tot} = \sum_i |\mu_i|$,
where $i$ runs over all images. 
{ 
The sign of the magnification reflects the parity of the image. That is, 
$\mu_i>0$ ($\mu_i<0$) corresponds to positive (negative) parity of the $i$-th
image, where the lens mapping is orientation-preserving (orientation-reversing). 
}
For derivations of these results, see e.g. \cite{Schneider92} and
\cite{Petters01}.

For the investigation in this paper, we derive the lens equation for a lensing system composed of
a thin circular belt with or without a point mass at the centre. We first 
show the face-on case to build an intuitive understanding and then generalize to
inclined belts. For simplicity, we assume the belt to have a 
uniform surface density. We choose the centre of the belt in projection as 
the origin of the source and lens planes, and in the case of the inclined belt, we set the directions of the belt's major 
and minor axes in the lens plane as $\bvec{\hat x}_I$ and $\bvec{\hat y}_I$, respectively.

\subsection{Face-on Systems}
\label{sec:face_on_sys}
As a first step, we consider a uniform thin ring seen face-on, 
centred at the origin, with radius $a$ (in units of the Einstein ring radius). 
This lensing geometry has been briefly studied before 
\citep[p.247]{Schneider92}, and here we present a more detailed analysis.

Rays passing outside the ring ($|r_I| > a$) experience a deflection as if the 
ring were a point mass located at the origin, while rays passing through the 
ring ($|r_I| < a$) are undeflected. The latter result can be 
easily inferred by 
considering two chords passing through the impact point with an infinitesimally small
opening angle between them -- 
the deflection angles caused by the two arc elements on the ring bounded by 
the two chords are the same but in the opposite directions, leading to null 
contribution to the deflection. Thus, the lensing equation reads
\be 
\bvr_S = \bvr_I  - \frac{\bvr_I}{|\bvr_I|^2}\Theta(|r_I| - a), 
\ee
where $\Theta(x)$ is the Heaviside step function. Although the lensing 
equation is written in a vector form, it reduces to a scalar equation
(by replacing $\bvr_S$ and $\bvr_I$ with $r_S$ and $r_I$, respectively)
given the symmetry of the face-on ring/belt case discussed in this subsection. 
For such a scalar equation, while the source position $r_S$ is positive by 
definition, the image position $r_I$ can be either positive or negative, 
corresponding to images located on the same side ($r_I > 0$) or 
opposite side ($r_I < 0$) of the lens as the source.

The generalization from a thin ring to a belt of finite width is straightforward. Only the 
mass inside the circle intersecting the impact point contributes to the deflection,
as if it were a point mass at the centre.
For a belt of constant surface density with inner and outer radii $a_i$ and 
$a_o$, we have
\be
\label{eq:belt_only}
\bvr_S = \bvr_I - \frac{\bvr_I}{|\bvr_I|^2} R\left(\frac{r_I^2-a_i^2}{a_o^2-a_i^2}\right),
\ee
where $R(x)$ is a modified ramp function defined as
\be
\label{eq:ramp}
R(x)=\left\{ 
        \begin{array}{ll}
        0 & \mbox{if } x\leq0,\\
        x & \mbox{if } 0<x\leq 1,\\
        1 & \mbox{if } x>1.
        \end{array}
\right. 
\ee

\begin{figure}
\includegraphics[width=\columnwidth]{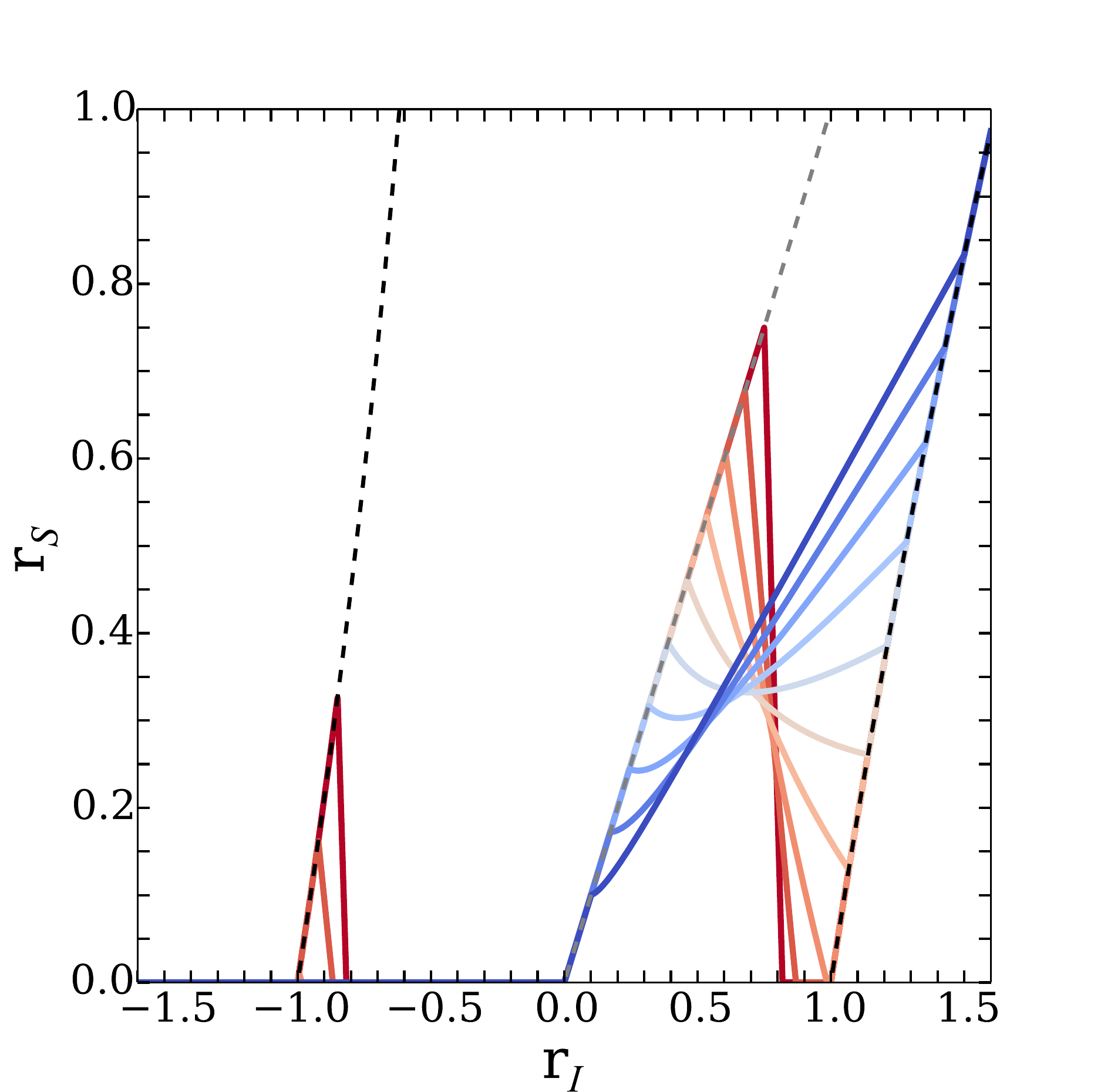}
\caption{
\label{fig:face_on_belt}
Solutions to the lens equation for a face-on belt with different widths. All 
belts have equal mass and each has a mean radius of $a_c = (a_i+a_o)/2 = 0.8$ 
(in units of the 
Einstein ring radius). The outer (black) dashed curves show the solution to the lensing equation if the belt were replaced by a point mass (with $r_S = r_I - 1/r_I$), while the inner (grey) dashed curve shows the solution if the belt had zero mass, and is simply given by $r_S = r_I$ (i.e. undeflected rays). 
The solid curves show the physically realized solutions, with the 
width of the belt increasing from red to blue. Note that in the limit of an 
infinitesimally thin ring the downward-sloping sections disappear (or become
vertical) and the corresponding images vanish, since they become zero-sized. Belts with high enough 
surface densities (red curves) can form 5 images, with two located on the opposite side of the lens as the source (with $r_I<0$)
}
\end{figure}

\begin{figure*}
\hspace{-2em}
\includegraphics[width=.85\columnwidth]{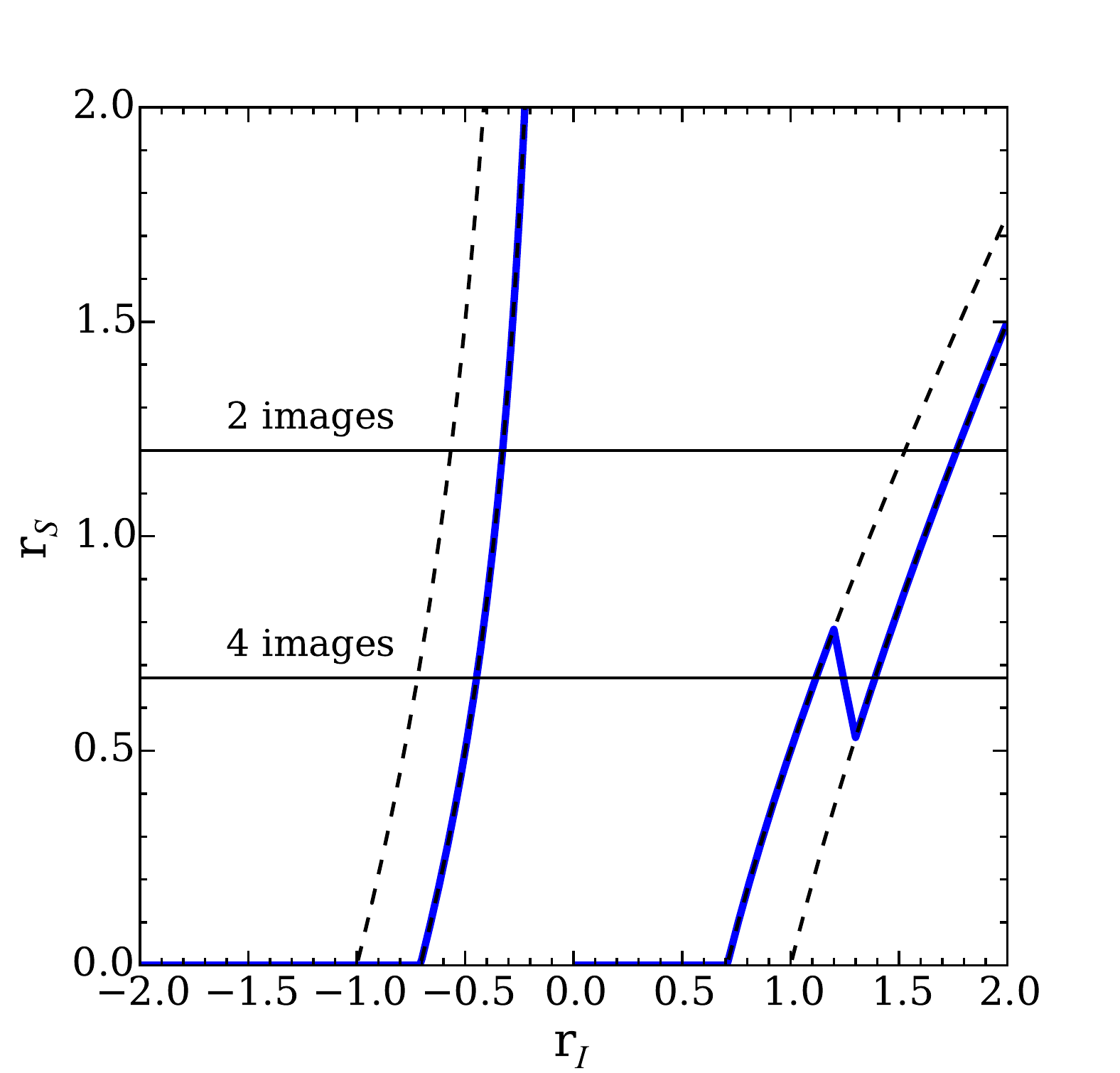}
\hspace{3em}
\includegraphics[width=.85\columnwidth]{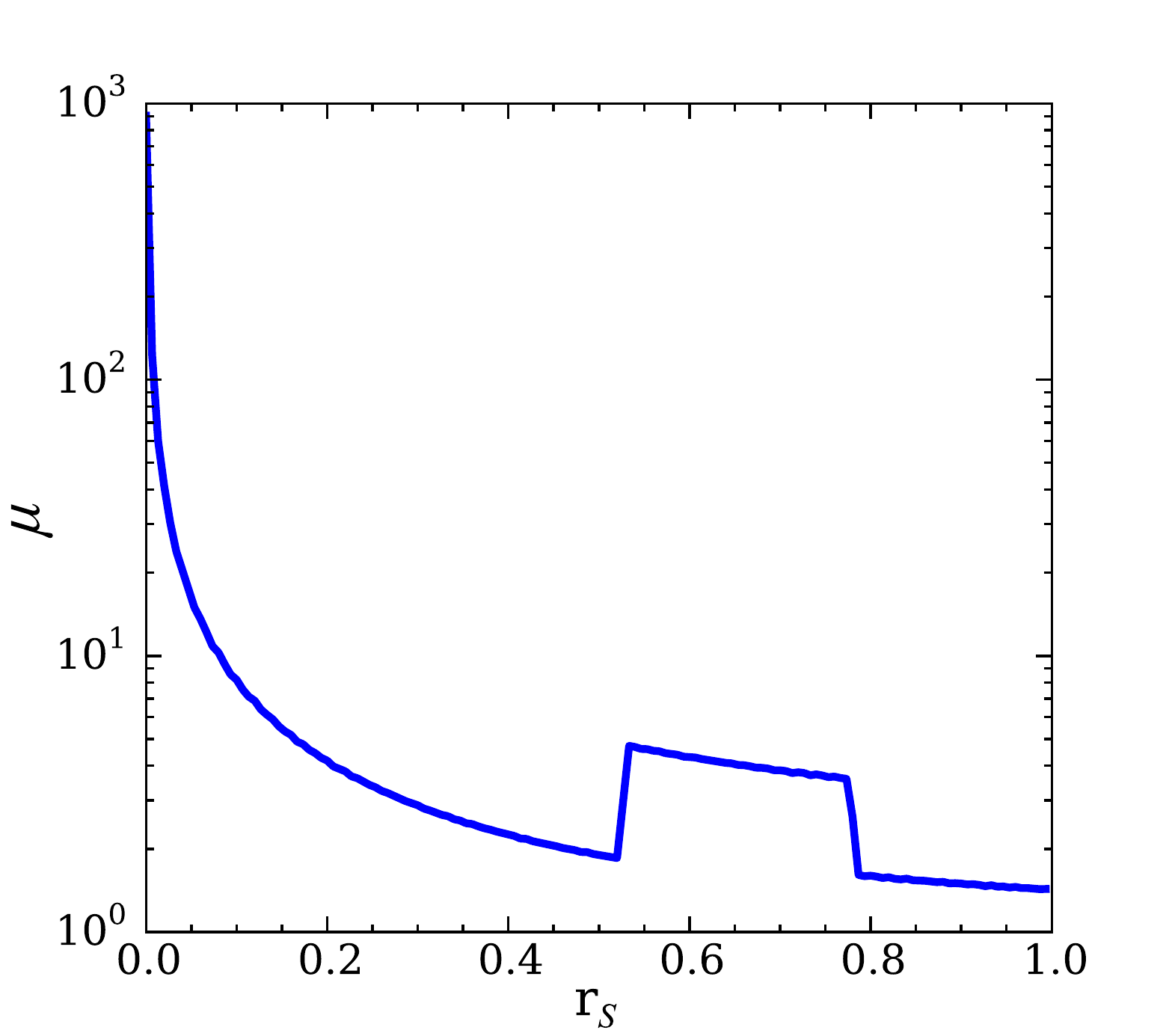}
\caption{
\label{fig:belt+point_soln}
{\it Left:} Solutions to the lens equation for a belt+point mass lens. The belt is 
face-on, with mean radius $a_c = 1.25$ and width $\Delta a = 0.1$. The belt to
point mass ratio is set to $q = 1.0$. The two inner (outer) black dashed curves 
correspond to the solutions from just the point mass (from the total belt+point 
mass system acting as a point mass). 
The blue solid curves are the physically realized solutions
for the belt+point mass case. The presence of the belt creates a region in the 
source plane where four images can be created. Negative $r_I$ corresponds to images located on the opposite
side of the lens as the source. 
{\it Right:} 
Magnification for the lens in the left plot as a function of the distance $r_S$ of 
the (point) source from the origin. The step-like jump in magnification marks 
the source positions corresponding to
the \pss\ (see the text), and the plateau bounded by the two steps corresponds 
to source positions where four images are formed. 
}
\end{figure*}

The solution to equation~(\ref{eq:belt_only}) is plotted in 
Figure~\ref{fig:face_on_belt} for a fixed belt mass with different values of 
the belt width $\Delta a = a_o - a_i$. 
With a mean radius $a_c = (a_i+a_o)/2=0.8$, the belt has width ranging from
$\Delta a = 0.1$ (dark red) to $\Delta a = 1.4$ (dark blue).
The middle dashed line corresponds to $\bvr_S = \bvr_I$, which is 
the solution for the undeflected rays passing interior to the inner edge of 
the belt ($|\bvr_I| < a_i$). The left and right dashed curves are the 
solution for rays deflected by the whole belt (i.e. equivalent to putting all 
the mass of the belt at the origin), which is the solution
for rays passing exterior to the outer edge of the belt ($|\bvr_I| > a_o$). 
We refer to the two solutions as the interior and exterior solution, 
respectively. For rays with $a_i < r_I < a_o$ that pass through
the belt, if the belt is wide (low surface density), 
$\partial r_S / \partial r_I >0$. However, if the belt is narrow (high surface 
density), $\partial r_S / \partial r_I <0$ for $a_i < r_I < a_o$. 
This is because as $r_I$
increases, the high surface density makes a substantial increase in the 
mass inside the radius of the impact point, which causes a large deflection 
and allows rays from small $r_S$ to be deflected to the observer.
The interior and exterior solutions are connected by rays passing through 
the belt, which complete the final solution. 
In Figure~\ref{fig:face_on_belt}, the number of images formed by a point 
source at a radius of $r_S$ in the source plane is found by counting the 
number of intersection points of the solution curve with a horizontal line 
at $r_S$. 

In the general case, the number of images that can form with a face-on belt
lens depends on the configuration of the belt. The lensing 
equation~(\ref{eq:belt_only}) is linear, quadratic, and quadratic for rays
passing interior to the inner edge of the belt, through the belt, and exterior 
to the outer edge of the belt, respectively. So we can have five solutions
in total. Since not all the solutions are physical given the above boundary 
conditions, we expect the maximum number of images formed by the 
belt to be $m_{max} = 5$

Let $m(\bvr_S)$ denote the number of images formed by a point source at a 
location $\bvr_S$ in the source plane. We find that in general, wide belts 
with low surface density have $m(\bvr_S) = m(r_S) = 1$ for all $r_S$, and 
$\partial r_S / \partial r_I > 0$ for all $r_I$. For narrow belts with high 
surface densities, we can have multiple images (either $m=3$ or $m = 5$) for small 
$r_S$, with $\partial r_S / \partial r_I <0$ for a finite range of 
$r_I$. The transition between these two regimes occurs at the critical belt 
width 
\be 
\label{eqn:delta_a}
(\Delta a)_{\rm cr} = \frac{2}{a_i + a_o} = \frac{1}{a_c}, 
\ee 
which is derived by setting $\partial r_S / \partial r_I |_{r_I = a_i^+} = 0$. 

In the case of $a_o > 1$, we find $m(r_S) = 1$ for small $r_S$ near the origin 
of the source plane. If $\Delta a > (\Delta a)_{\rm cr}$, $m(r_S) = 1$ for all 
$r_S$, as $r_S$ is a strictly monotonically increasing function of $r_I$. 
If $\Delta a < (\Delta a)_{\rm cr}$, the belt lens leads to an annulus in the 
source plane where $m=3$. In the limit of the thin ring, the central image 
(from rays passing through the belt) disappears, since its size is zero.
The remaining two images correspond to undeflected rays passing interior to 
the ring and rays passing exterior to the ring that are then deflected to the 
observer. 

If $a_o < 1$, we always have $\Delta a < (\Delta a)_{\rm cr}$, and as a result 
$m(r_S) > 1$ at some point in the source plane. Intuitively, if $a_o < 1$, 
the entire mass of the belt lies within its own Einstein radius (with the 
belt shrinking to a point mass as the extreme case), and so the belt can form 
more than one image for small $r_S$.
We find that $m=5$ near the origin in the source plane and $m=3$ in an annular 
region surrounding the central $m=5$ region, which becomes an $m=1$ region at 
large $r_S$. 

All the above features can be understood precisely by considering the behavior 
of the total source-plane magnification $\mu(r_S)$ of the lens mapping. Based on the solution 
of the belt-only case, we see that as a source moves from $r_S = 0$ to $r_S \gg 1$, 
$\mu(r_S)$ changes discontinuously
as the images (or impact points) cross the belt. The change is caused by the 
discontinuity in $\partial r_S / \partial r_I$ at $|r_I| = a_i$ or $a_o$, 
which corresponds to a change in the number of images $m$ for a narrow belt 
with $\Delta a < (\Delta a)_{\rm cr}$ or a sharp change in the image size for a 
wide belt with $\Delta a > (\Delta a)_{\rm cr}$. The loci in the source plane 
associated with the discontinuous change in $\mu(r_S)$ are defined as 
pseudo-caustics, which we will define more precisely in \S 3. 
Since jumps in $\mu(r_S)$ occur when the image of a source crosses the inner 
or outer edge of the belt, the \pss\ can be found simply by 
mapping circles with radius $a_i$ and $a_o$ in the lens (image) plane into the 
source plane. From this and equation~(\ref{eq:belt_only}), we see that these 
lensing systems always possess two concentric circular pseudo-caustics, 
located at source radii 
\be 
\label{eqn:beltonly_pss} 
\mathcal{PS}_1 = a_i \quad {\rm and} \quad \mathcal{PS}_2 = \left|a_o - \frac{1}{a_o} \right|. 
\ee
For belts with $\Delta a < (\Delta a)_{\rm cr}$, which can form regions with 
$m>1$, the source radii $r_S = \mathcal{PS}_1$ and $r_S=\mathcal{PS}_2$ 
define the boundaries of the different image multiplicity regions mentioned 
previously. 

We now introduce a point mass to our lensing system, located at the centre 
of the face-on belt. We denote the mass ratio of the belt to the point mass 
as $q = M_{\rm belt}/M_{\rm point}$ and normalize the angular positions by 
the the Einstein ring radius for the combined belt+point system. This has the 
effect of adding an extra term (the lensing contribution from the point mass) 
to the lens equation, with equation~(\ref{eq:belt_only}) becoming
\be
\label{eqn:faceon_belt_point}
\bvr_S=\bvr_I-\frac{1}{1+q}\frac{\bvr_I}{|\bvr_I|^2} - \frac{q}{1+q}\frac{\bvr_I}{|\bvr_I|^2}R\left(\frac{r_I^2-a_i^2}{a_o^2-a_i^2}\right).
\ee

Just like the belt-only lens, the belt+point mass lens possesses pseudo-caustics which correspond to the inner and outer boundaries of the belt mapped into the source plane. In this case, the pseudo-caustics take the form of two concentric rings at source-plane radii of 
\be 
\label{eqn:ps_pointbelt}
\mathcal{PS}_1 = \left| a_i - \frac{1}{(1+q)a_i}\right| \quad {\rm and} \quad \mathcal{PS}_2 = \left|a_o - \frac{1}{a_o}\right|. 
\ee
We can also derive the condition for pseudo-caustics that change the image multiplicity, as opposed to those that just change the size of images. As before, this is done by finding the parameters at which $r_S$ ceases to become a monotonically increasing function of $r_I$. Given an inner radius $a_i$, we denote $(a_o)_{\rm cr}$ as the maximum outer radius of the belt that allows for image-forming pseudo-caustics. We obtain 
\be 
\ba 
& a_{o,{\rm cr}}  = \sqrt{\frac{2q}{1+q+1/a_i^2} + a_i^2}, \quad \text{if $a_{o,{\rm cr}} > 1$}; \\ 
& a_{o,{\rm cr}}  = \frac{1}{\sqrt{2q+2}} \bigg[ a_i^2(1+q) + q-1   + \bigg(a_i^4(1+q)^2 \\ & \qquad\qquad + (q-1)^2 + 2a_i^2q(3q^2+4q+1) \bigg)^{1/2} \bigg]^{1/2},\quad\text{if $a_{o,{\rm cr}} \leq 1$}. 
\ea
\ee
In the belt-only limit ($q\rightarrow +\infty$), it can be shown that the 
above equations reduce to equation~(\ref{eqn:delta_a}).

The solution to equation~(\ref{eqn:faceon_belt_point}) is plotted in the left panel of 
Figure~\ref{fig:belt+point_soln} for a face-on belt with 
$a_i = 1.2,~a_o=1.3,$ and $q = 1$. 
The two middle curves (with the left one solid blue and the right one blue transitioning to dashed black) 
correspond to the possible solutions to the lensing
equation for rays passing interior to the belt, where only the central point
mass is responsible for the deflection. The two dashed curves at large 
$|r_I|$
are the possible solutions for rays passing exterior to the belt, where the 
deflection comes from the total mass of the system.
The solutions include images located at both positive and negative $r_I$, which correspond 
to images with positive and negative parities, and are located on the same side of the lens as the source and on the opposite side, respectively. 
The far left dashed curve includes rays deflected from 
both the belt and the point mass, but the rays in fact pass interior to the belt
(i.e. $|r_I|<a_i$), meaning that this solution is not physically realized.
The middle right dashed curve corresponds to rays passing interior to the
belt and deflected only by the point mass. For $|r_I|>a_i$, deflection from 
the belt starts to contribute and so this solution is not allowed. Similarly,
for the rightmost dashed curve, the allowed solution should have an image 
position beyond the outer radius of the belt, i.e. $|r_I|>a_o$. 
As a whole, the solid blue curves represent the physically realized solution. The 
transition from the interior solution to the exterior solution occurs when the 
images lie within the range of the belt, i.e. when $a_i<|r_I|<a_o$. Since the 
belt is narrow and $a_o < a_{o,{\rm cr}}$, we have $\partial r_S / \partial r_I < 0$ 
in this part of the solution (see Figure~\ref{fig:face_on_belt} for a 
comparison).

The right panel of Figure~\ref{fig:belt+point_soln} shows the total 
magnification a point source experiences when lensed by the face-on belt+point 
system, as a function of distance $r_S$ from the origin in the source plane. 
At both small and large $r_S$, $m=2$ (see the left panel of 
Figure~\ref{fig:belt+point_soln}), while in the region bounded by the
two \pss\ (see equation~(\ref{eqn:ps_pointbelt}); which corresponds to 
$69/130 \leq r_S \leq 47/60$ for the case here), we find $m=4$. 
In the magnification map, as the number of images changes
from two to four, there is a sharp increase in the magnification (about a 
factor of 2.5). Note that in this case, the source does not need to cross a formal caustic (defined as 
a curve in the source plane where $|\mu(\bvr_S)| = \infty$ for a point source)
for the number of images to change. The associated loci across which the 
magnification is finite but changes discontinuously are the \pss mentioned in the introduction. The existence of \pss\ implies
that we cannot rely solely on the formal caustics in order to understand 
these lensing systems. 

\subsection{Inclined Systems}\label{sec:inclined_systems}

We now allow our system to be inclined, with the normal of the plane of the belt inclined at an 
angle $i$ relative to the line of sight. We again start with the case of a
uniform ring with radius $a$, which becomes an ellipse when seen in 
projection. The ellipse is parametrized by the set of vectors in the
lens (image) plane,
which satisfy 
\be
\frac{x^2}{a^2} + \frac{y^2}{b^2} = 1,
\ee
where $a$ is the ring's semi-major axis and $b = a\cos i$ is the ring's semi-minor 
axis. Even for arbitrary inclinations, rays passing interior to the ring are remarkably 
still undeflected \citep{Chandrasekhar69,Bray84}. 

To find the deflection 
for an impact point ${\bvec{r}_I}$ in the image plane 
that lies outside the ring (i.e. the 
exterior solution), we consider 
an ellipse passing through ${\bvec{r}_I}$ and confocal to the projected ring. 
The confocal ellipse is expressed as
\be \label{eq:confocal_ellipse}
\frac{x^2}{a'^2} + \frac{y^2}{b'^2} = 1,
\ee
where the primed semi-axes are defined by $a'^2 = a^2 + \lambda$ and 
$b'^2 = b^2 + \lambda$, with $\lambda$ found by substituting 
$\bvec{r}_I=(x_I,y_I)$ into equation~(\ref{eq:confocal_ellipse}) and taking the positive solution.

The deflection angle $\bvec{\alpha}_r({\bvec{r}_I}) = \alpha_{1}\bvec{\hat x}_I 
+ \alpha_{2} \bvec{\hat y}_I$ from the ring can be found analytically using 
results from ellipsoidal potential theory \citep{Bourassa75}, which are solved nicely in 
\cite{Schramm90}. For $\bvr_I = x_I\bvec{\hat x}_I + y_I\bvec{\hat y}_I$, the deflection angle from the ring is 
\be \label{eqn:alpha_ring}
\bvec{\alpha}_r(\bvec{r}_I;a,i) = p'^2\left(\frac{\bvec{x}_I}{a'^3b'} + \frac{\bvec{y}_I}{a'b'^3}\right),
\ee
where $p' = \left( x_I^2/a'^4 + y_I^2/b'^4 \right)^{-1/2}$ is the perpendicular
distance from the centre of the confocal ellipse to the tangent line at 
$\bvec{r}_I$ and we also use the notation $\bvec{x}_I=x_I\bvec{\hat x}_I$ and $\bvec{y}_I=y_I\bvec{\hat y}_I$ for simplicity.

\begin{figure*}
\includegraphics[width=.75\textwidth]{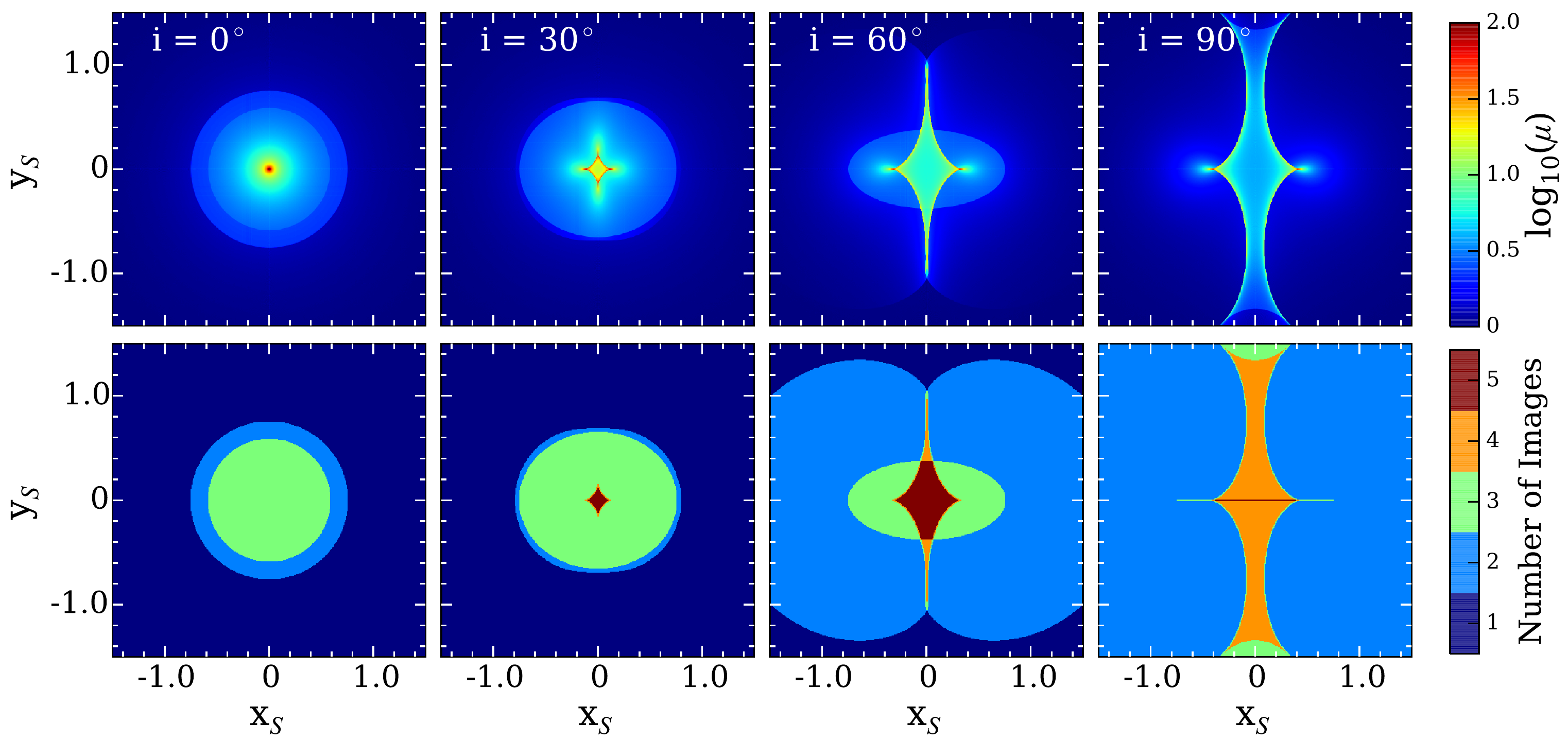}
\includegraphics[width=.75\textwidth]{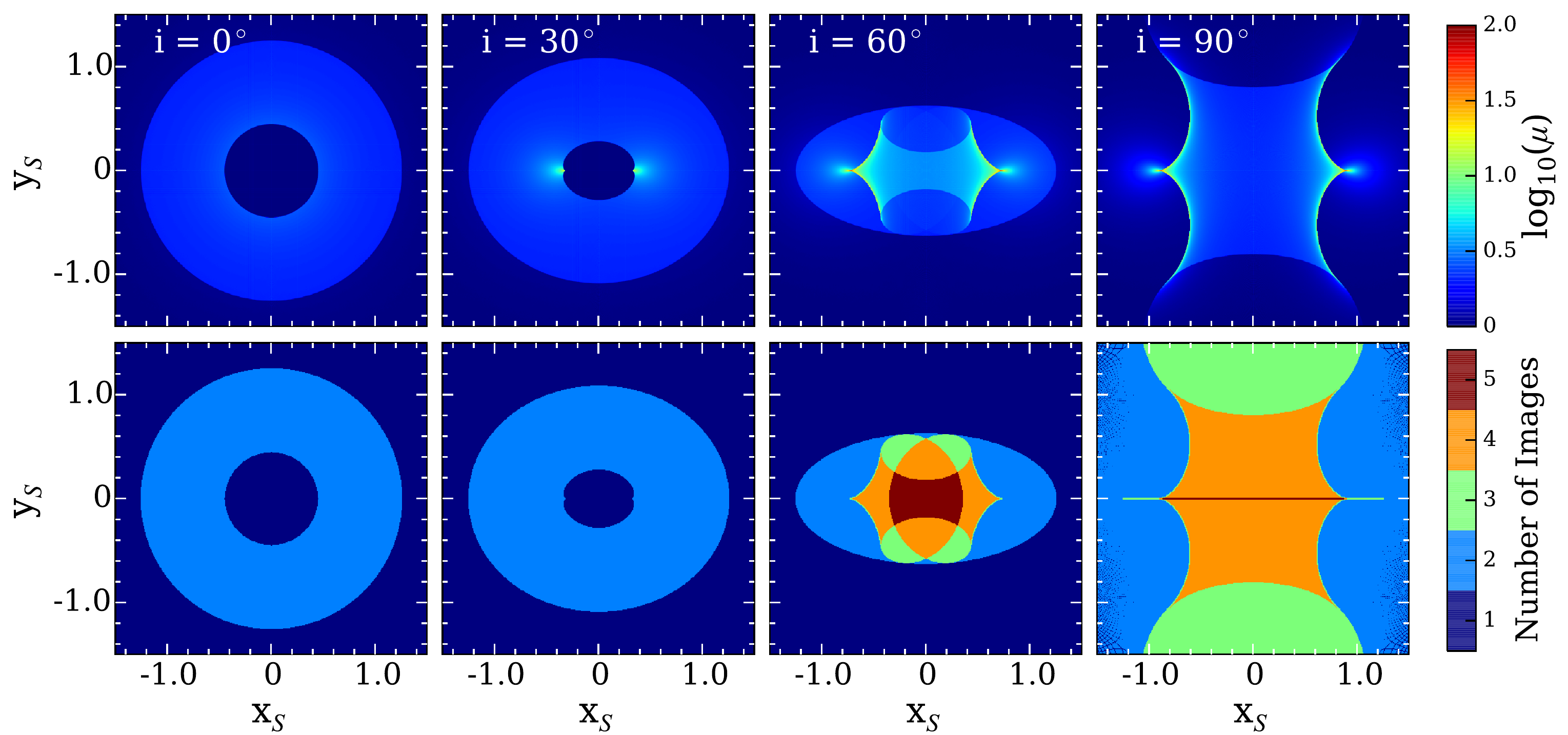}
\caption{
\label{fig:8panel_ringonly}
Magnification and image multiplicity maps for a ring-only lens. The top plot 
shows a ring of radius $a = 0.75$ across a range of inclinations. The top 
row shows the magnification $\mu$ in the source plane, with the bottom row 
showing the number of images a source at a given position produces. The bottom 
plot is the same as the upper plot, but with the ring radius changed to 
$a = 1.25$. Note that in both cases there are loci with finite change in
magnification and with corresponding change in the number of images, which are
\pss\ discussed in the text.
}
\end{figure*}

To obtain an expression for the deflection caused by a belt, we make use of 
more results in ellipsoidal potential theory. Consider a circular uniform
disc in projection, which appears as an elliptical disc with 
constant normalized surface density $\kappa_0$, semi-major axis $a$, and inclination $i$. 
The normalized deflection angle at a point $\bvec{r_I}$ in the image plane
is \citep{Schramm90}
\be \label{eq:belt_alpha}
\bvec{\alpha}_d(\bvr_I;a,i) = 2\pi \tilde{a} \tilde{b} \kappa_0 \left( \frac{1}{a'+b'}\frac{\bvec{x}_I}{a'} +\frac{1}{a'+b'}\frac{\bvec{y}_I}{b'}\right), 
\ee
where for points interior to the disc, we 
set $a'^2 = \tilde{a}^2 = x_I^2 + y_I^2 / \cos^2 i$ and $b' = \tilde{b} = a'\cos i$. For 
points exterior to the disk, we set $\tilde{a} = a$ and $\tilde{b} = b$ and obtain $a'$ and $b'$ by substituting $(x_I,y_I)$ into equation~(\ref{eq:confocal_ellipse}). 

To calculate the 
deflection angle of 
a belt of inner and outer semi-major axes $a_i$ and $a_o$, we superimpose 
a disc with semi-major axis $a_o$, inclination $i$, and surface density $\kappa_0$ with a 
disc of effective surface density $-\kappa_0$, semi-major axis $a_i$, and inclination $i$. 
The total deflection from the two superimposed 
discs is just that from a uniform belt
with inner and outer radii $a_i$ and $a_o$:
\be \bvec{\alpha}_b(\bvr_I;a_i,a_o,i) = \bvec{\alpha}_d(\bvr_I;a_o,i) - \bvec{\alpha}_d(\bvr_I;a_i,i). \ee
Here, $\kappa_0$ is 
defined as $\kappa_0 = 1/[\pi(a_ob_o - a_ib_i)]$ for the belt so that the total 
dimensionless mass of the belt is unity. It is easy to verify that 
the belt does not deflect rays that pass within 
the inner edge of the belt.

With the deflection angle from the belt, the lens
equation for the general case of an inclined belt+point system then becomes
\be
\bvec{r}_S=\bvec{r}_I - \frac{1}{1+q} \frac{\bvr_I}{|\bvr_I|^2} 
-\frac{q}{1+q}\bvec{\alpha}_b(\bvr_I;a_i,a_o,i), 
\ee
where $q$ is the belt-to-point mass ratio. The second term on the right-hand
side corresponds to the deflection from the centre point mass. Taking 
the limit of $q\rightarrow +\infty$ recovers the equation for the belt-only case. 

While computing the deflection for systems with arbitrary inclination is straightforward, analytically solving the lensing 
equation becomes much more cumbersome. However, it is possible to get a relatively simple analytic classification for the 
two limiting cases of the inclination $i$ -- we have addressed the face-on $i=0$ case in the previous section, and present 
an analysis of the edge-on $i=\pi/2$ case in Appendix~\ref{sec:appendix_edgeon}. 

Equipped with the solutions of the deflection angle and the lens equation
for the inclined system and using our understanding of the face-on system, 
we turn now to an investigation of the general lensing properties of belt+point mass systems
as a function of the inclination.

\section{Lensing Properties of Belt+Point Mass Systems: Caustics and Pseudo-Caustics}

To study the lensing properties of belt+point lensing systems, we employ the 
inverse ray-shooting technique
\citep[e.g.][]{Schneider86}. Light rays are drawn uniformly from
the image plane and mapped back to the source plane according to the lens
equation. The magnification at each position in the source plane is computed
as the ratio of the density of rays at this angular position to that in the 
image
plane, which is also the density of rays in the source plane in the absence of 
the lens.

Of interest in analysing the properties of lensing systems is the behaviour of 
caustic curves, which are loci in the source plane where the Jacobian 
matrix of the lens mapping is singular (${\rm det}(A)=0$), and where the 
magnification $\mu(\bvr_S)$ is formally infinite. 
{ 
The caustic curves correspond to the singular points of the lens mapping. 
For smooth lens mappings, we may invoke Sard's theorem to show that the set of caustic
curves must have measure zero, which means that they
must consist of collections of one-dimensional curves and isolated points.
}

For all but the most simple lensing geometries, finding analytical 
expressions for the caustic curves is difficult, since the mapping 
$\bvec{r}_I \mapsto \bvec{r_S}$ is highly non-linear. While the inverse-ray 
shooting technique enables us to find high magnification regions, 
it is limited by the resolution of the mapping. To better locate 
the caustic curves, we use equation~(\ref{eq:detmag}) to track down points 
where the magnification is above some threshold value (we adopt $|\mu(\bvr_S)| > 1000$).
In Appendix~\ref{sec:analytic_mu_appendix}, we provide analytic expressions for the magnification 
in each of the systems we consider.

{ 
For isolated lensing systems\footnote{A lensing system is {\it isolated} if the time delay function $T_{\bvr_S}(\bvr_I) = \frac{1}{2}|\bvr_I-\bvr_S|^2 - \psi(\bvr_I)$, which is the sum of the geometric and gravitational time delays, 
satisfies $T_{\bvr_S}(\bvr_I) \rightarrow \infty$ as $|\bvr_I| \rightarrow \infty$ for all finite, non-caustic $\bvr_S$.} with lensing potentials $\psi(\bvr_I)$ that are smooth for all $\bvr_I$ 
away from a finite number of point singularities, several results from singularity theory 
can be applied to constrain the types of images that sources can form.
In particular, the image multiplicity $m(\bvr_S)$ can only ever change by two. This 
can be seen by applying the Poincare-Hopf index theorem, which states that the sum 
$\sum_i {\rm sgn}(\mu_i)$ is a constant for all non-caustic source locations $\bvr_S$, 
and implies that images can only be created / annihilated in opposite-parity 
pairs \citep{Burke81,McKenzie85}. For mass distributions $\kappa(\bvr_I)$
with no point singularities, this implies that 
the image multiplicity $m(\bvr_S)$ must be odd for all $\bvr_S$, since we expect a source at $\bvr_S$ to 
form only one, undeflected image with positive parity 
in the limit $|\bvr_S| \rightarrow \infty$. 
More generally, if $\psi(\bvr_I)$ possesses $g$ point singularities, the total image multiplicity is always odd (even) if 
$g$ is even (odd), with a source at $|\bvr_S| \rightarrow \infty$ forming one undeflected positive-parity image
and $g$ negative-parity images located at the positions of each singularity \citep{Petters01}. 

Additionally, for isolated lenses with smooth $\psi(\bvr_I)$, the image multiplicity $m(\bvr_S)$ 
can be constrained to change when, 
and only when, the source crosses a caustic \citep{Schneider92}. This can be seen by considering
the map $\bvr_S \rightarrow m(\bvr_S)$, which associates each point $\bvr_S$ in the source plane
with the total number of images formed by a point source located at $\bvr_S$. Since $m(\bvr_S)$ is an integer it cannot 
change continuously, and hence for smooth lens mappings $m(\bvr_S)$ can 
only change when the source crosses a caustic. 
Because of this, the caustic set must consist of {\it closed} curves (together 
with possible isolated point caustics). 
Indeed, if there were open caustic curves, a closed loop could be constructed in the source plane 
that had only one intersection with a caustic curve. A source traveling 
along this loop would then arrive back at its starting point with a different 
image multiplicity than it started with, which is impossible. 

These constraints on image formation can be relaxed in systems, like the ones considered in this paper, whose lensing potential $\psi(\bvr_I)$ 
satisfies weaker differentiability or singularity conditions. Most relevantly for us, these constraints 
can be relaxed if $\psi(\bvr_I)$ is not twice-differentiable ($C^2$). 
By Poisson's equation $\nabla^2_{\bvr_I}\psi(\bvr_I) = 2\kappa(\bvr_I)$,
this is guaranteed if $\kappa(\bvr_I)$ is discontinuous, 
which occurs for the majority of the 
lensing geometries considered in this paper (although we consider the case of a $C^\infty$ mass distribution in Sec. \ref{sec:smooth_distributions}).  
As was already shown in the face-on ring case and will be studied in more 
detail below, discontinuous mass distributions may allow for the existence of 
regions in the source plane across which the image multiplicity can change 
by one, violating the invariance of the image parity.

In the literature, curves in the source plane across which the image multiplicity changes 
by one are often called ``pseudo-caustics'' (or ``cuts'' in e.g. \citealt{Kovner87}). Several previous
studies have dealt with pseudo-caustics, primarily in the study of variants of singular isothermal spheres and ellipsoids \citep[e.g.][]{Wang97,Evans98,Keeton00,Rhie10},
with which the ring+point lens shares several features. In most of these models, the \pss\ appear 
as large smooth circles or ellipses. For example, the singular isothermal sphere model
is defined by the lensing potential $\psi(\bvr_I) = R|\bvr_I|$ (implying $\kappa(\bvr_I) \propto 1/|\bvr_I|$), and possesses a circular \ps\ at $r_S = R$,
where the number of images changes by one. As a source traveling from $r_S < R$ to 
$r_S > R$ crosses the \ps, a negative-parity image ``disappears'' into the singularity at 
$r_I = 0$. This is similar to what happens to the images of sources crossing the 
circular \pss\ in the face-on ring examples considered earlier, when a negative-parity image
``disappears behind the ring'' and becomes zero-sized.

In this paper, we adopt a slightly more general definition of \pss, which is better 
suited for discussing the lensing systems we consider. We define \pss\ to be 
curves in the source plane across which the magnification $\mu(\bvr_S)$ changes 
discontinuously, and yet remains finite. This definition is more general than 
the definition in terms of image multiplicities --- besides including loci in 
the source plane where the number of images changes by one, it also includes 
curves across which the size of an image jumps sharply,
which will be relevant in what follows. 

The \pss\ are found to possess a broad variety of morphologies across the 
parameter space of the lensing systems we consider. 
The formal caustics correspond to the degeneracy of the lensing Jacobian
(implying $\mu(\bvr_S) \rightarrow \infty$), and as a consequence they are 
usually cusped. By contrast, the \pss\ are defined to always have finite $\mu$,
and they always appear to be smooth.
In many cases, we will see that the existence of \pss\ can 
lead to the presence of open caustics. 
}

It is tempting to attribute the presence of \pss\ in the thin ring case to the 
discontinuity of the lens mapping between the inner-ring and outer-ring solutions. However, the 
belt lensing systems we consider also exhibit a wide variety of \ps\ features, 
even though the deflection angle $\bvec{\alpha}_b(\bvr_I;a_i,a_o,i)$ is continuous throughout the entire lens plane. Rather, we will see that it is the 
{\em non-differentiability} of the deflection angle (and hence the discontinuity of $\kappa(\bvr_I)$)
that causes the formation of the \pss. 

Since the \pss\ represent loci where $\mu$ is 
non-differentiable, they can in principle be found by finding points in the 
source plane across which the magnification is discontinuous. 
However, the explicit analytic expression for $|\nabla_{\bvr_S} \mu|$ is very 
unwieldy. In practice, we find the \pss\ by using our inverse ray-shooting results to 
numerically compute the magnification map in the source plane and the 
associated image multiplicity map, which then allows us to find the pseudo-caustics by inspection.

{ 
As was the case for face-on rings and belts, we will see that the \pss\ correspond to the boundaries of the belt/ring in the
image plane mapped to the source plane, which allows us to derive an analytic expression for the \pss\ (see Appendix 
\ref{sec:pss_expression}). Thus, for the models we consider, we can associate discontinuities 
in $\kappa(\bvr_I)$ with discontinuities in $\mu(\bvr_S)$, which define the \pss. We demonstrate this 
by explicitly mapping the belt/ring boundaries into the source plane, and checking that they correspond to regions 
in which $\mu(\bvr_S)$ is finite and discontinuous. We note that this is not a general 
proof that discontinuities in $\kappa(\bvr_I)$
are in a one-to-one correspondence with discontinuities in $\mu(\bvr_S)$, 
and that this correspondence may not hold for other types of lensing systems. 
}

\subsection{Ring-Only Systems}
\label{sec:ring-only}

To proceed, we first turn our attention to the simplest case -- that of a 
ring-only lensing system. Figure~\ref{fig:8panel_ringonly} shows magnification 
and image multiplicity maps for a range of ring parameters.

The top plot in Figure~\ref{fig:8panel_ringonly} shows a ring with radius $a = 0.75$, while in the bottom plot the ring 
has $a = 1.25$. In both cases, the inclination of the ring increases from 
left to right, going from face-on to edge-on. The top rows of each plot show 
magnification, while the bottom panels indicate the number of images for
a source, $m(\bvr_S)$, at the given location in the source plane. We compute $m(\bvr_S)$ by sweeping over the image plane, identifying the set of all points
$\{(x_I,y_I)\}$ that map to within a circle of radius $\epsilon \ll 1$ centred at $\bvr_S$, and then 
count the number of path-connected regions in the set $\{(x_I,y_I)\}$. 

In the $a = 0.75$ case (top plot), we see many boundaries between regions with different
values of $m$, where the magnification has a finite jump. 
These are the \ps\ curves mentioned above. In the face-on case (leftmost panel),
from small to large radii, we see a circular $m=3$ region, an annular 
$m=2$ region, and finally the $m=1$ region, divided by the two 
\ps\ curves. Examining our earlier results on face-on systems explains that the ring 
can have $m = 3$ for small $\bvr_S$ because $a < 1$. Applying equation~(\ref{eqn:beltonly_pss})
in the limit $a_i = a_o = a$ allows us to derive the radii 
of the \ps\ circles as $\mathcal{PS}_1 = a = 0.75$ and 
$\mathcal{PS}_2 = a^{-1} - a \approx 0.58 $.
The different image multiplicity regions can also be understood by plotting the solutions to the lens
equation in a way similar to Figures~\ref{fig:face_on_belt} and 
\ref{fig:belt+point_soln}. There are three possible solutions, one 
(from $\bvr_S=\bvr_I$) corresponding to the undeflected light passing interior to the 
ring (the interior solution) and the other two (from $\bvr_S=\bvr_I(1-1/|\bvr_I|^2)$) corresponding to the deflection 
by a central point with the same mass as the ring (the exterior solutions). 
The physically realized 
solutions are determined by the ring radius. For example, at $|\bvr_S|\gg 1$ only
one of the exterior solutions is physical, giving rise to the $m=1$ region.

The solution for the $m=3$ region has one image from rays passing 
interior to the ring (hence undeflected), so the corresponding region 
in the image plane must be interior to the 
ring, which can be used to 
track the inclination of the ring. Indeed, as the ring becomes more inclined,
the region becomes more elliptical (top plot of 
Fig.\ref{fig:8panel_ringonly}). As the ring inclines, a $m=5$ diamond-shaped
region forms near the centre and a vertical structure of increased 
magnification appears. Since the projected mass of the inclined ring becomes more and 
more concentrated toward the two ends of the major axis, the normal caustic 
feature exhibits some similarity to that from an equal-mass binary lensing system 
with the two point masses symmetrically placed on the major axis, with the
positions approaching $\pm a$ as the ring becomes edge-on.
That is, as the inclination increases, the caustic features mimic 
those of close- and intermediate-separation equal-mass binary point mass
\citep[e.g.][]{Schneider86,Liebig15} and binary isothermal sphere \citep{Shin08} lenses. 
{ Additionally, the combination of a large circular \ps\ and a central asteroidal caustic 
seen in the $i=30^{\circ}$ panel is similar to the caustic structures of singular isothermal
sphere potentials and variants thereof (see e.g. Figure 1 of \citealt{Rhie10} 
and Figure 5 of \citealt{Kovner87}).  }

The case with $a = 1.25$ (bottom plot) is fairly different. When seen face-on, 
the ring is no longer concentrated enough to form multiple images near the 
centre, since the entire mass distribution of the ring lies outside of its own Einstein radius. 
The possible physical solution to the lens equation includes that
from the undeflected rays and that corresponding to the case of a point-mass 
lens (the one with the image outside the Einstein ring radius), with the 
former and latter applying to small and large radii, respectively. In between, 
there is an annular region where both apply, and the boundaries give the two 
\ps\ curves, which are circles of radii $\mathcal{PS}_1= a = 1.25$ and 
$\mathcal{PS}_2 = a - a^{-1} = 0.45 $.
As the ring becomes inclined, this annular region pinches down about the major
axis, and two folded regions of high magnification appear at the inner parts 
of the annulus. As the inclination continues to increase, regions with higher 
image multiplicities start to form. As with the $a = 0.75$ case, the normal 
caustics share similarities with those from an intermediate-separation 
equal-mass binary lens. For both values of $a$, we see the appearance of a narrow
high-$m$ region located along the $x_S$ axis. For a point source the additional images formed 
in this region occur when the $y_S$-coordinate of the source is strictly zero, and 
so the corresponding high-$m$ region has zero thickness. 

We now turn our attention to the ring+point mass system and provide an illustrative
understanding of the \pss\ and open caustics. Finally, we present the results 
for the belt+point mass system.

\subsection{Ring/Belt+Point Mass Systems}

\begin{figure*}
\vspace{-10pt}
\includegraphics[width=.75\textwidth]{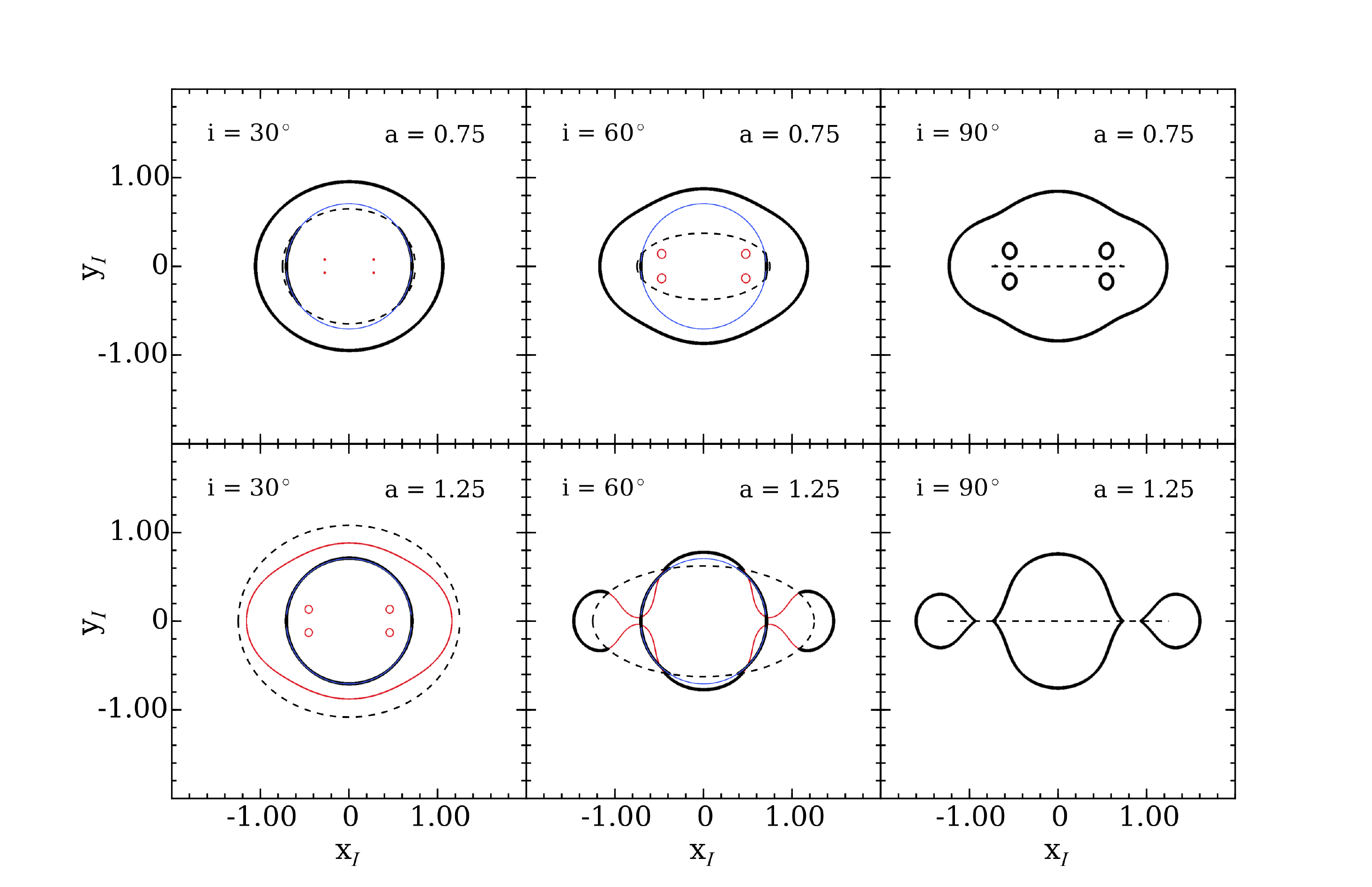}
\includegraphics[width=.75\textwidth]{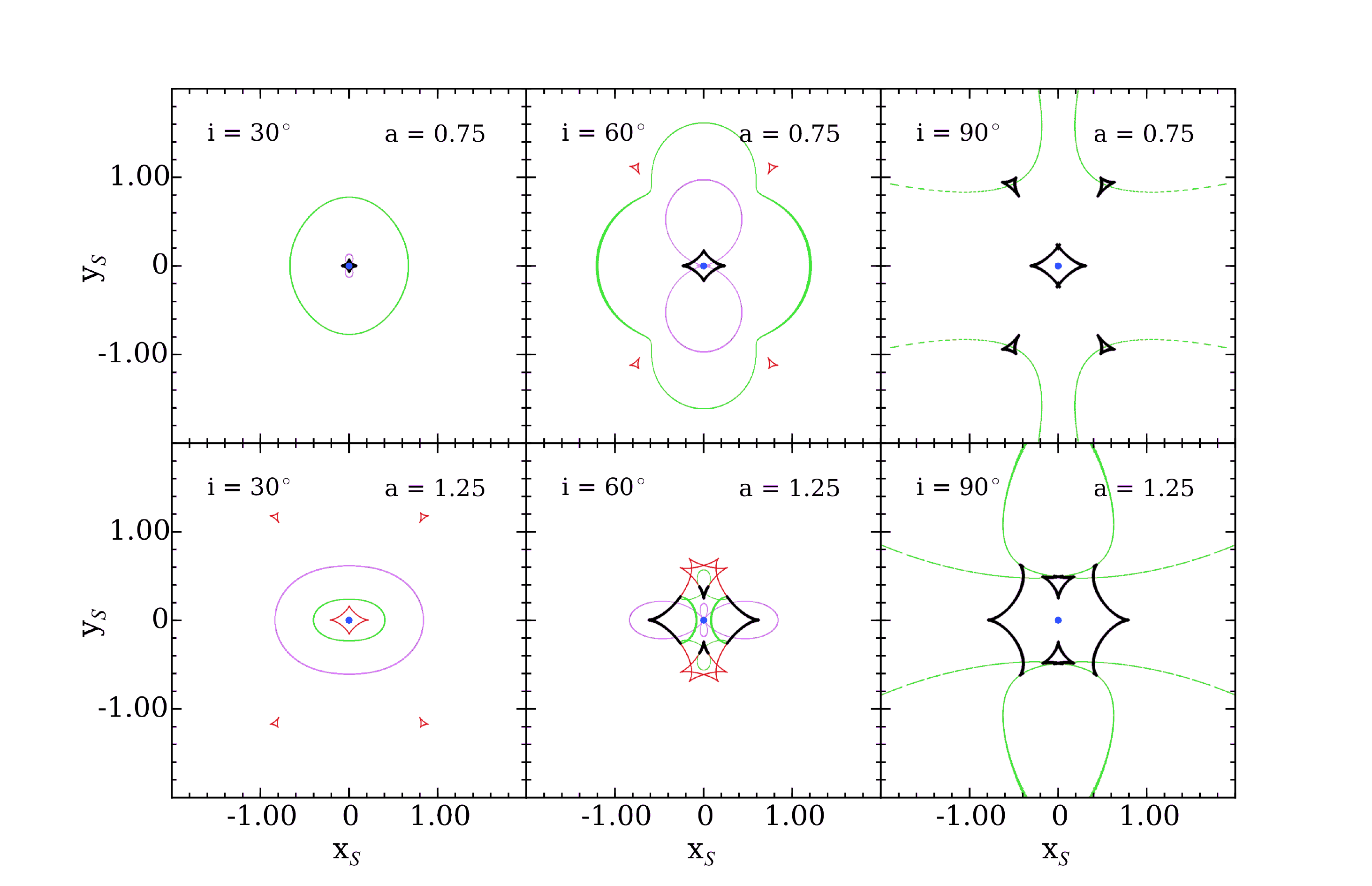}
\caption{
\label{fig:caustics_plus_criticals}
Critical curves (top plot) and caustic curves (bottom plot) for a ring+point 
lens. The ring and the point mass have equal masses. The radius and inclination
of the ring are labelled in each panel. In each critical-curve panel, the
blue (red) thin curves connected with the thick solid ones form the ``full''
critical curves from the interior (exterior) solution. The dashed curve 
delineates the projected ring in the image (lens) plane. The critical curves 
from the interior (exterior) solution becomes invalid for the parts
outside (inside) the ring. The final allowed critical curves are in thick 
black. The corresponding caustics are denoted with the same colour and line
types in the caustic-curve panels. The ring is mapped onto the source plane
according to the interior (purple) and exterior (green) solution, 
respectively. These mapped ring curves form the pseudo-caustics.
See the text for detail.
}
\end{figure*}

\subsubsection{Ring+Point Mass System}

\begin{figure*}
\includegraphics[width=.75\textwidth]{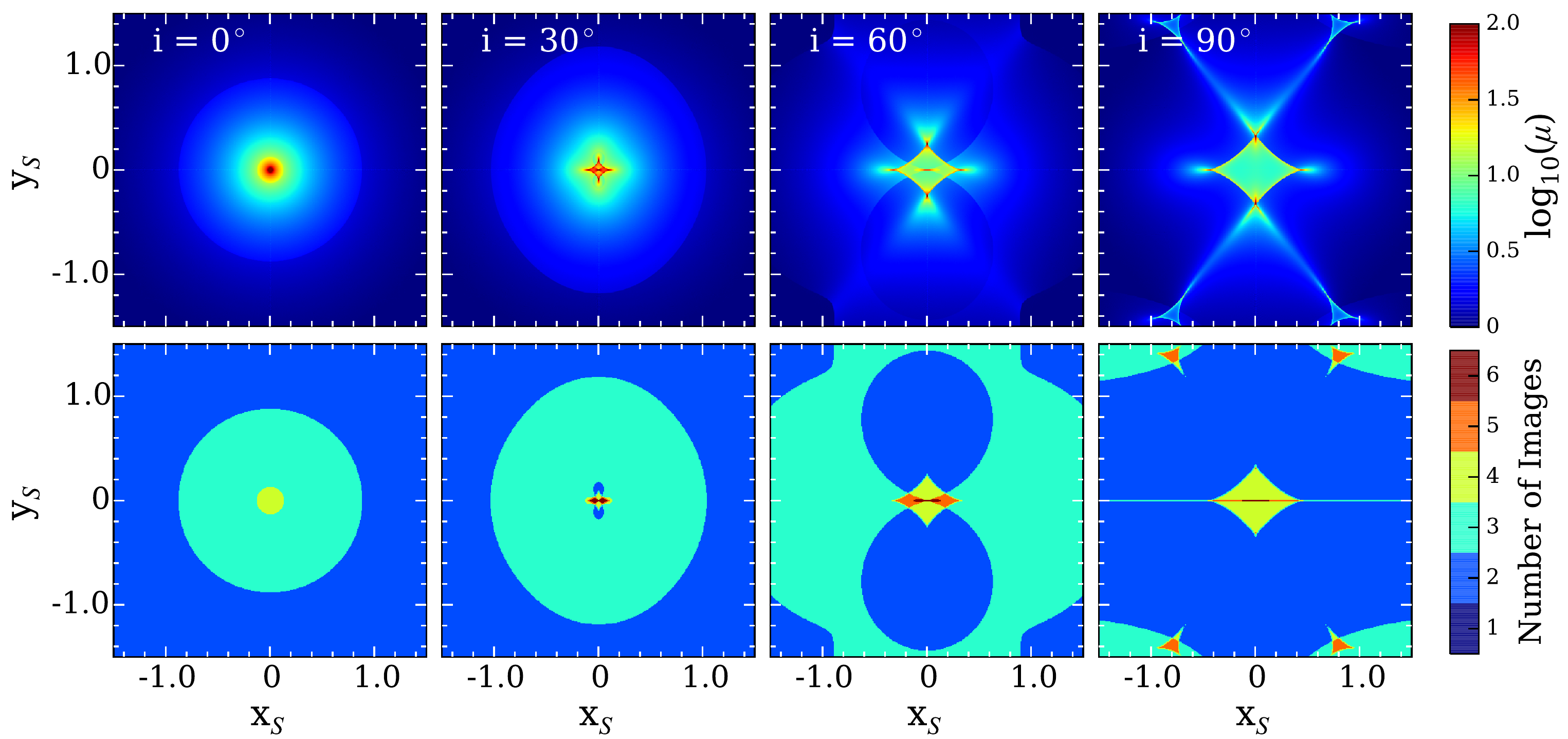}
\includegraphics[width=.75\textwidth]{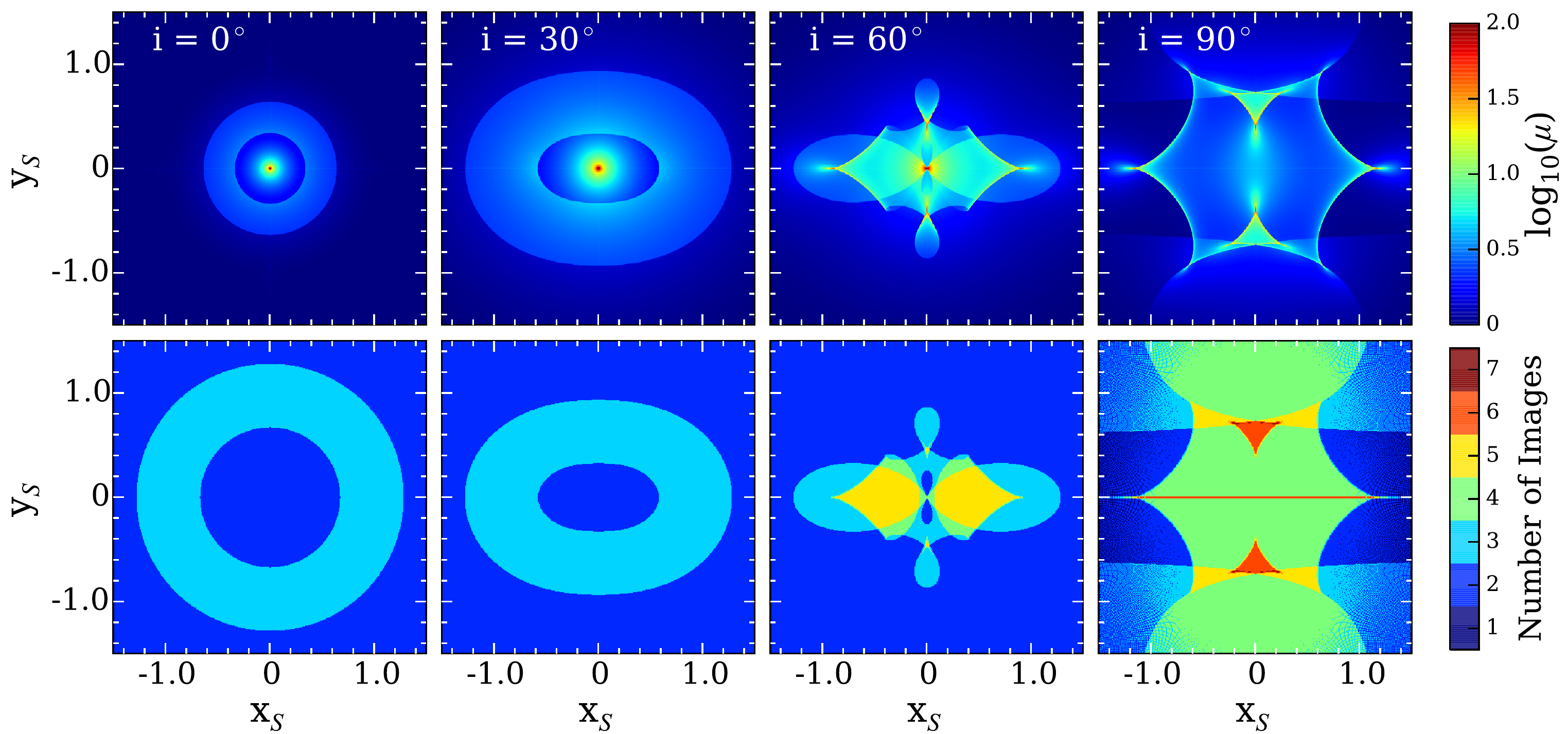}
\caption{
\label{fig:8panel_ringstar}
Magnification and image multiplicity maps for a ring+point lens. The ring and 
point mass have equal masses. The ring has radius $a = 0.75$ in the top plot and
$a = 1.25$ in the bottom plot. 
As in Figure~\ref{fig:8panel_ringonly} there exist \pss\, associated with 
changes in the image multiplicity.
See Figure~\ref{fig:caustics_plus_criticals} for the corresponding \pss\
associated with the cases with $i\geqslant 30^\circ$.
}
\end{figure*}

With a point mass placed at the centre of the ring/belt, the solutions 
become more complex. Examining the critical curves in the 
image plane and the corresponding caustic curves in the source plane is
helpful for us to understand the lensing properties, especially the
\pss\ and open caustics.

The thick black curves in Figure~\ref{fig:caustics_plus_criticals} 
are the critical curves (top plots) and the corresponding caustic curves
(bottom plots) for the ring+point mass system. To understand these curves,
it is useful to present the {\it full} sets of critical and caustic curves 
for the interior and exterior solutions, which are shown as the blue and 
red curves, respectively. For the interior solution, 
the term ``full'' refers to the curves determined from the
centre point mass, as if the ring did not exist. For the exterior solution,
``full'' means that we plainly apply the exterior solution even if the rays
pass interior to the ring. Or effectively, we shrink the projected ring
{\it confocally} to a degenerate ellipse with zero minor axis so that the
exterior solution can be extended to small radii. We investigate 
which parts of the critical curves are allowed to form the final critical
curves (thick black) given the configuration of the system, and deduce 
the \pss\ and open caustics. For the system shown in 
Figure~\ref{fig:caustics_plus_criticals}, the ring and point 
mass have equal masses, with the ring having a radius of either 
$a = 0.75$ or $a = 1.25$ as labelled in the panels. The inclination of 
the ring increases from $30^{\circ}$ to $90^{\circ}$ from left to right.
The scales are normalized to the Einstein ring radius of the 
combined ring+point mass system.

The ``full'' critical curve for the interior solution is simply a circle with 
radius $1/\sqrt{2}$, the Einstein ring for the centre point mass (since 
we normalize by the Einstein ring radius for the total 
ring+point mass, which is twice the point mass here). The 
corresponding caustic is the point at the origin. For the exterior solution, 
the ``full'' critical curve is also a circle (the Einstein ring for the total 
ring+point mass) for 
the system at face-on. As the ring becomes more inclined, its mass 
distribution approaches more toward a binary lens with the two equal point 
masses symmetrically placed on the $x$-axis and on the two sides of the 
centre point mass toward $(\pm a,0)$. The whole system mimics the one composed 
of three point masses \citep{Danek15}. For the $a=0.75$ case, besides a big ring-like 
critical curve, four small ringlet-like critical curves are formed. The big
and small curves correspond to the central diamond-shaped caustic and the
four planetary cusp caustics, respectively. As a whole, at high inclination,
the critical and caustic curves share similarities with the close-separation 
triple point lensing system. For the $a=1.25$ case, the analysis is similar and at 
high inclination we find the system to be similar to the intermediate- or 
wide-separation triple point lens. 

With the ``full'' critical and caustic curves in place for both the interior 
and exterior solutions, we now analyse which parts of the curves are physically realized
by considering the validity of the solutions. Unlike in 
Section~\ref{sec:ring-only} where we explain things in the source plane,
we find it is more intuitive to work in the image plane, since the validity of the solutions is determined by the image 
position (whether it is interior or exterior to the ring) and since the projected 
ring is also naturally described in the image plane.

The dashed black curve in each critical-curve panel of 
Figure~\ref{fig:caustics_plus_criticals} delineates the projected ring.
For the interior solution, images that form outside of the projected ring are not physical. 
Therefore the part of the critical curve outside the ring (indicated by thin blue lines)
is not allowed. The allowed lens-plane area for the interior solution becomes smaller 
as the ring becomes more inclined. The corresponding caustic is still the point 
at the origin of the source plane.

For the exterior solution, the situation is reversed. The part of the
critical curve inside the ring (indicated by thin red
lines) is not allowed. In a few cases (e.g. with $a=0.75$ and $i=60^\circ$), the four 
small ringlet-like critical curves are inside the ring and hence are not
allowed, which leads to the disappearance of the ``planetary'' caustics. For
the case with $a=1.25$ and $i=60^\circ$, the allowed part of the critical 
curve forms four separated arclets, as a result of the truncation imposed 
by the ring. They lead to four separated caustic curves, and the truncation 
in the critical curve makes the caustic curve become open.

Since it is the projected position of the ring that determines which solution is valid,
the \pss\ can be obtained by mapping the projected 
ring onto the source plane, according to the interior and exterior solution, 
respectively. The inner (outer) boundary of the mapped ring is shown as purple (green) curves in the 
source 
plane that result from mapping the projected ring with the interior (exterior) solution in the caustic-curve 
panels of Figure~\ref{fig:caustics_plus_criticals}. Formal caustics can 
become open as they intersect the mapped ring. Since the mapping 
properties change discontinuously as the impact point (image) crosses the ring, the curves 
of the mapped ring are just the \pss\ mentioned before. Notice that each 
\ps\ is smooth, since the lens mapping is continuous just within and just outside
the ring (which also holds for the thick belt case). Since the magnification is finite
at the \pss, magnification and image multiplicity maps must be constructed
to study them.

Figure~\ref{fig:8panel_ringstar} shows the magnification and image multiplicity 
maps for the ring+point systems shown in 
Figure~\ref{fig:caustics_plus_criticals}, with $q=1$.
Clearly, the main features in the magnification map correspond closely 
with the formal caustics and \pss\ in Figure~\ref{fig:caustics_plus_criticals}.

For $a=0.75$ (top plot), when seen face-on, we see two
pseudo-caustics, similar to the ring-only case. 
However, the regions inside the \pss\ have higher image multiplicities than 
the ring-only case. The reason is that the presence of the centre point mass
allows the maximum number of solutions to be four, opposed to three as in the ring-only
case. Using our earlier results for face-on belt+point mass systems and taking $a_i = a_o = a$,
we can derive the radii of the \ps\ circles to be 
$\mathcal{PS}_1 \approx 0.11$ and $\mathcal{PS}_2 \approx 0.82$.
As the inclination increases, the three-image region becomes larger 
and more oblate. Interestingly, two lobes of {\it decreased} image 
multiplicity appear within the $m=3$ region. When seen edge-on, these 
regions become larger and push out the $m=3$ regions to larger radii. 

In the bottom plot of Figure~\ref{fig:8panel_ringstar}, the ring is enlarged 
to have a semi-major axis of $a = 1.25$. The \pss\ in this case appear distinct 
from the $a = 0.75$ case. When seen face-on, we see an $m=3$ annular 
region bounded by the two \pss. When the ring inclines, the annulus folds down 
over itself, creating complicated overlapping features (e.g. in the 
panel with $i=60^\circ$) from the interplay between the \pss\ and the 
formal caustics. When seen edge-on, $m=1$ regions and small $m=7$ 
regions are formed. Although edge-on $a = 0.75$ and $a = 1.25$ ring+point mass 
systems are very similar geometrically, they approximately correspond to 
close- and intermediate-separation triple point mass lensing systems and hence the 
magnification maps are different. 

\subsubsection{Belt+Point Mass Systems}

We now widen the ring to a belt, and study the resulting effects on the 
morphology of the \pss\ and the general magnification pattern. The lensing solutions are affected 
by the projected mass density (i.e. the width of the belt), as seen
in Figure~\ref{fig:face_on_belt}. For the following examples, we fix the belt width to be 
$\Delta a = 0.3$. In the belt+point mass case, the \pss\ correspond to the
inner and outer edges of the belt mapped onto the source plane according to
the interior and exterior solutions, respectively. The explicit expression
for the \pss\ can be found in Appendix~\ref{sec:pss_expression}.

Figure~\ref{fig:8panel_beltstar} shows the magnification and image 
multiplicity maps for belts with different mean radii. With
$\Delta a = 0.3$, the belt radius is centred at $a_c = 0.75$ in the top plot 
and $a_c = 1.25$ in the middle plot. In the bottom plot it is situated so that 
half of its mass lies at $a > 1$ and the other half lies at $a < 1$. In all 
plots, we take $q=1$.

\begin{figure*}
\includegraphics[width=.75\textwidth]{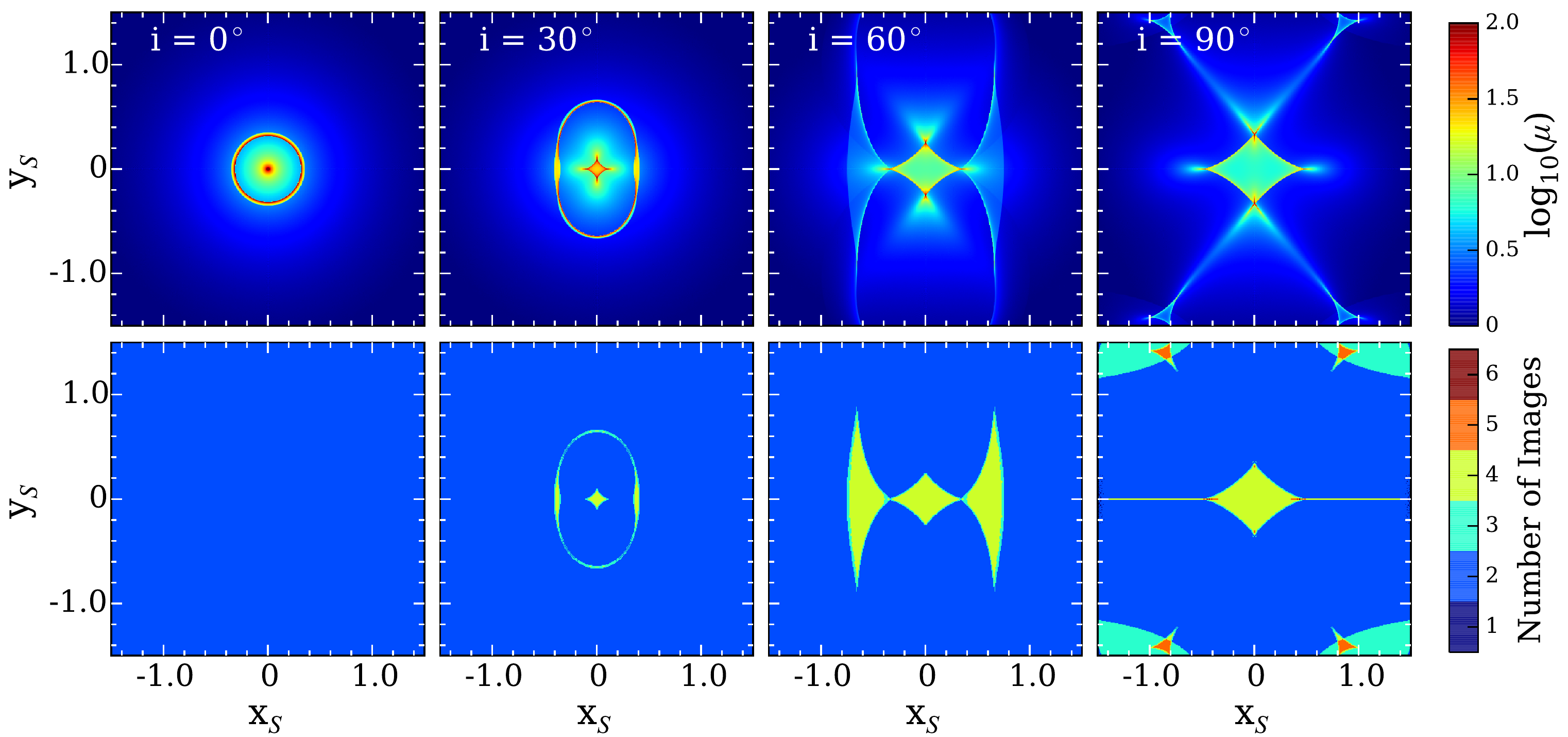}
\includegraphics[width=.75\textwidth]{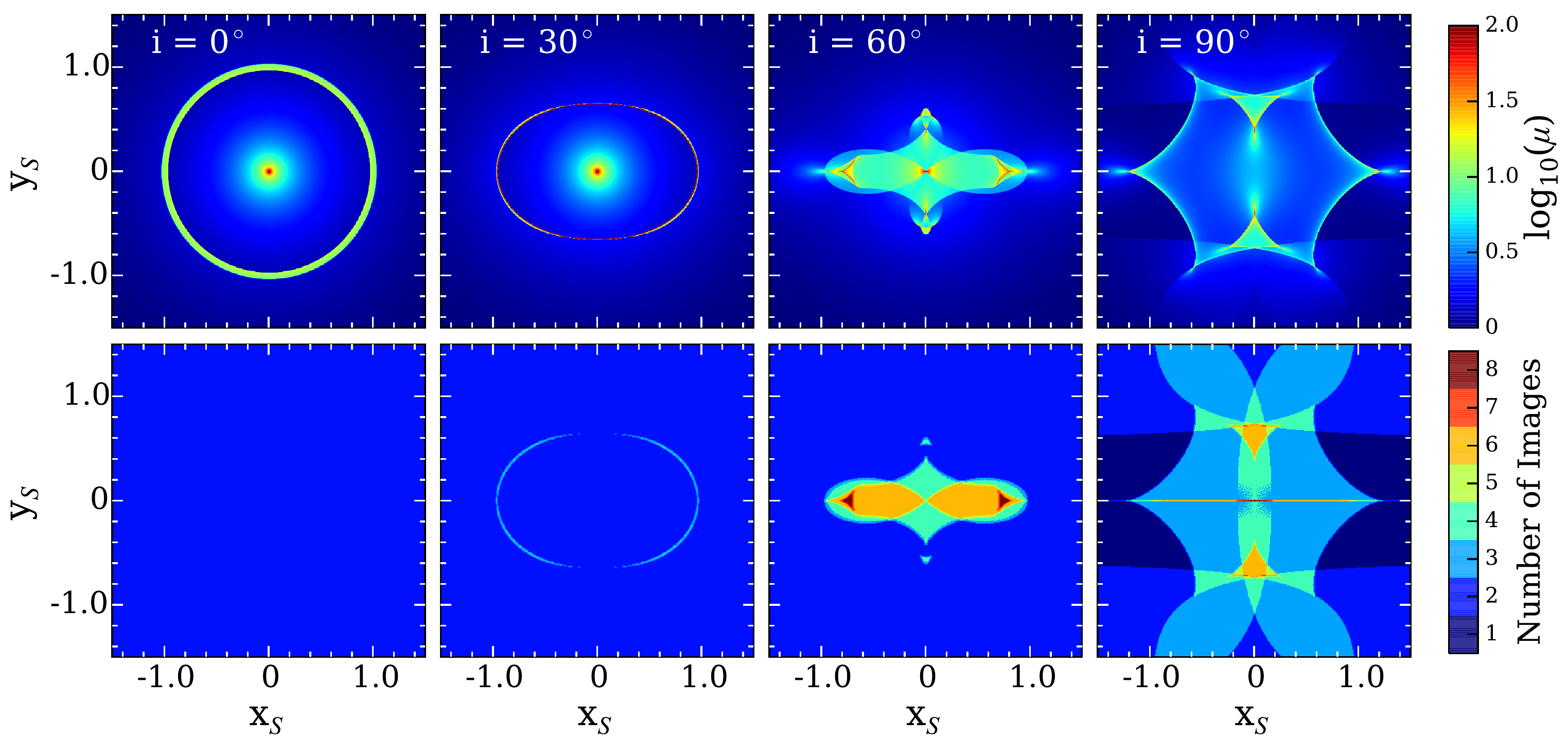}
\includegraphics[width=.75\textwidth]{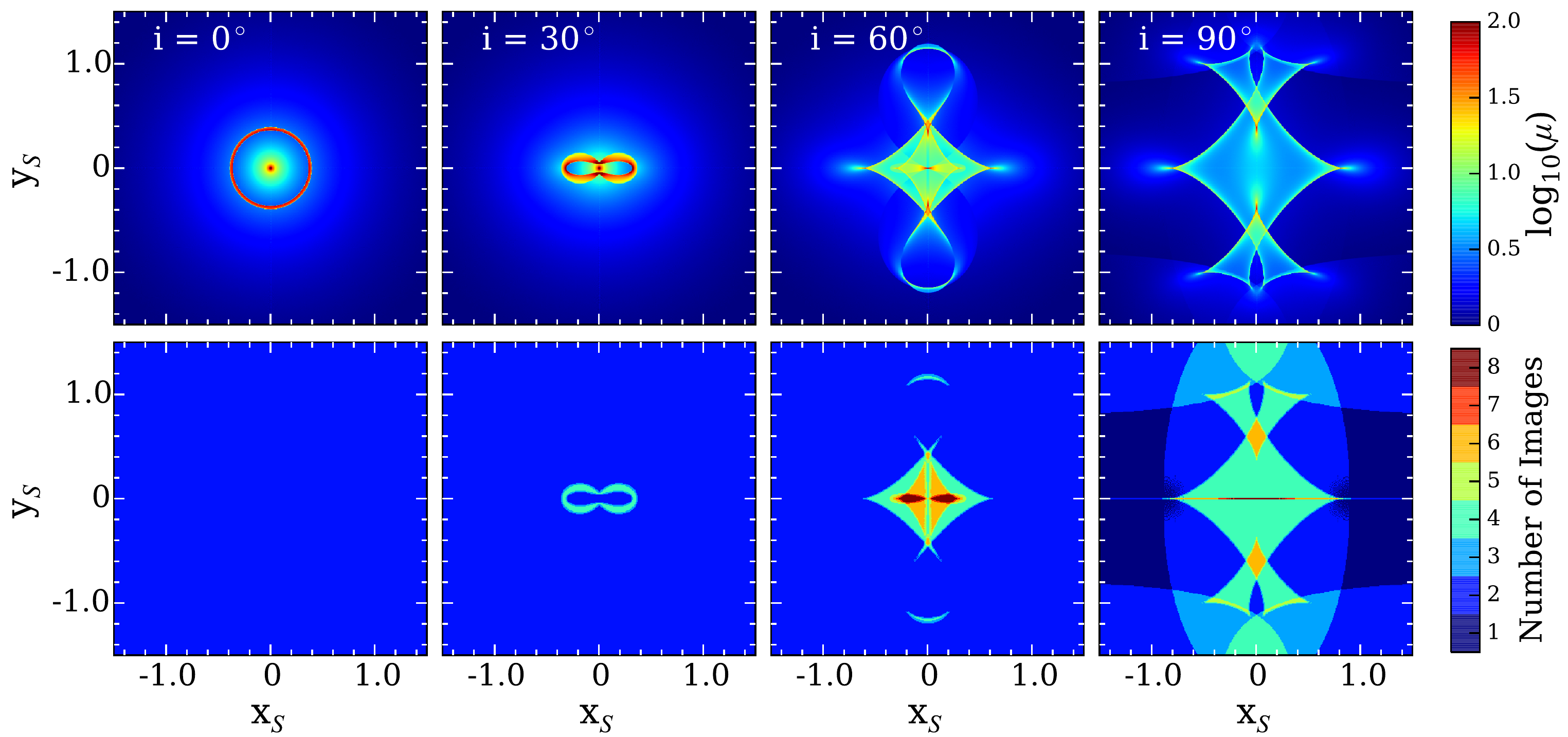}
\caption{
\label{fig:8panel_beltstar}
Magnification and image multiplicity maps for a belt+point mass lens at different
belt inclinations. 
In each plot, the belt has a width of $\Delta a = 0.3$. In the top plot, the 
radius of the belt is centred at $a_c = 0.75$, in the middle plot it is 
centred at $a_c = 1.25$, and in the bottom plot the belt is situated so that 
half of its mass is at $a < 1$ and the other half is at $a > 1$. 
In all the cases, there exist \pss\ associated with a discontinuous
change in magnification and a corresponding change in the number or size 
of images.
}
\end{figure*}

For $a=0.75$, we only see two images formed for the
face-on case since $a_o < a_{o,{\rm cr}}$, and the surface density of the belt is not high enough to 
generate additional images, similar to the case indicated by the blue 
curve in Figure~\ref{fig:face_on_belt}. However, we see a ring-like 
structure present in the magnification map, with no correspondent image 
multiplicity change. This ring-like structure corresponds to rays passing 
through the belt, experiencing deflection from both the belt and the centre point mass. 
This leads to the change in the slope of
the $r_S$-$r_I$ relation (as illustrated in Figure~\ref{fig:face_on_belt}).
This means that one of the images becomes larger in size, resulting in an overall increase
in magnification ($dr_I^2/dr_S^2$). This example shows that \pss\ need not to 
be associated with changes in the image multiplicity -- they can also 
correspond to discontinuous changes in the size of an image. 

When the belt inclines, a narrow $m=4$ annular region bounded by \pss\ 
appears that evolves into a fan-shaped region at $i = 60^{\circ}$. 
When the belt is seen edge-on, the magnification map is similar to 
the case of the $a = 0.75$ edge-on ring (right panels in the top plot of 
Figure~\ref{fig:8panel_ringstar}), as both approximately approach the 
close-separated triple point lensing system.

\begin{figure*}
\includegraphics[width=.55\textwidth]{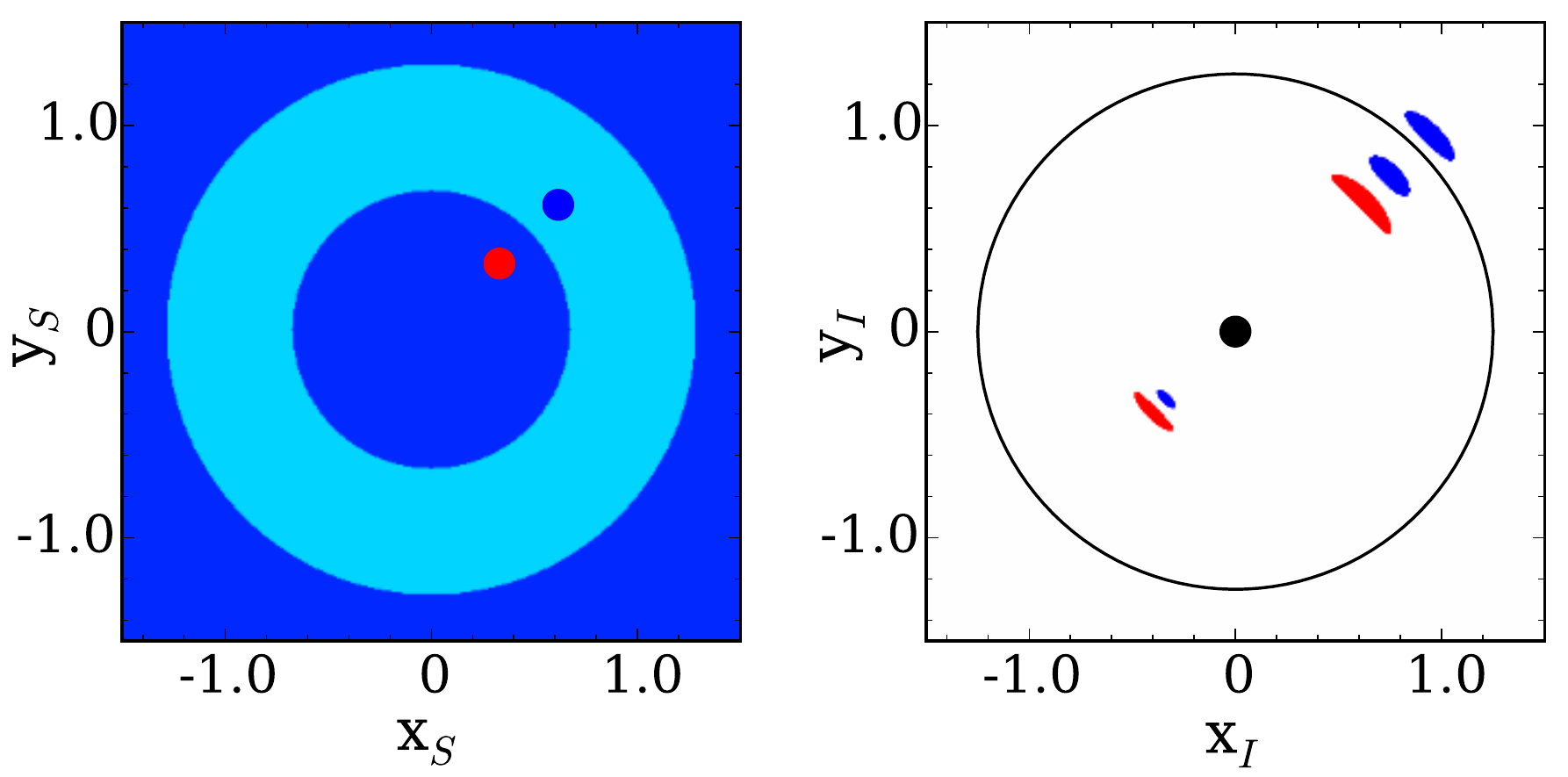}
\includegraphics[width=.55\textwidth]{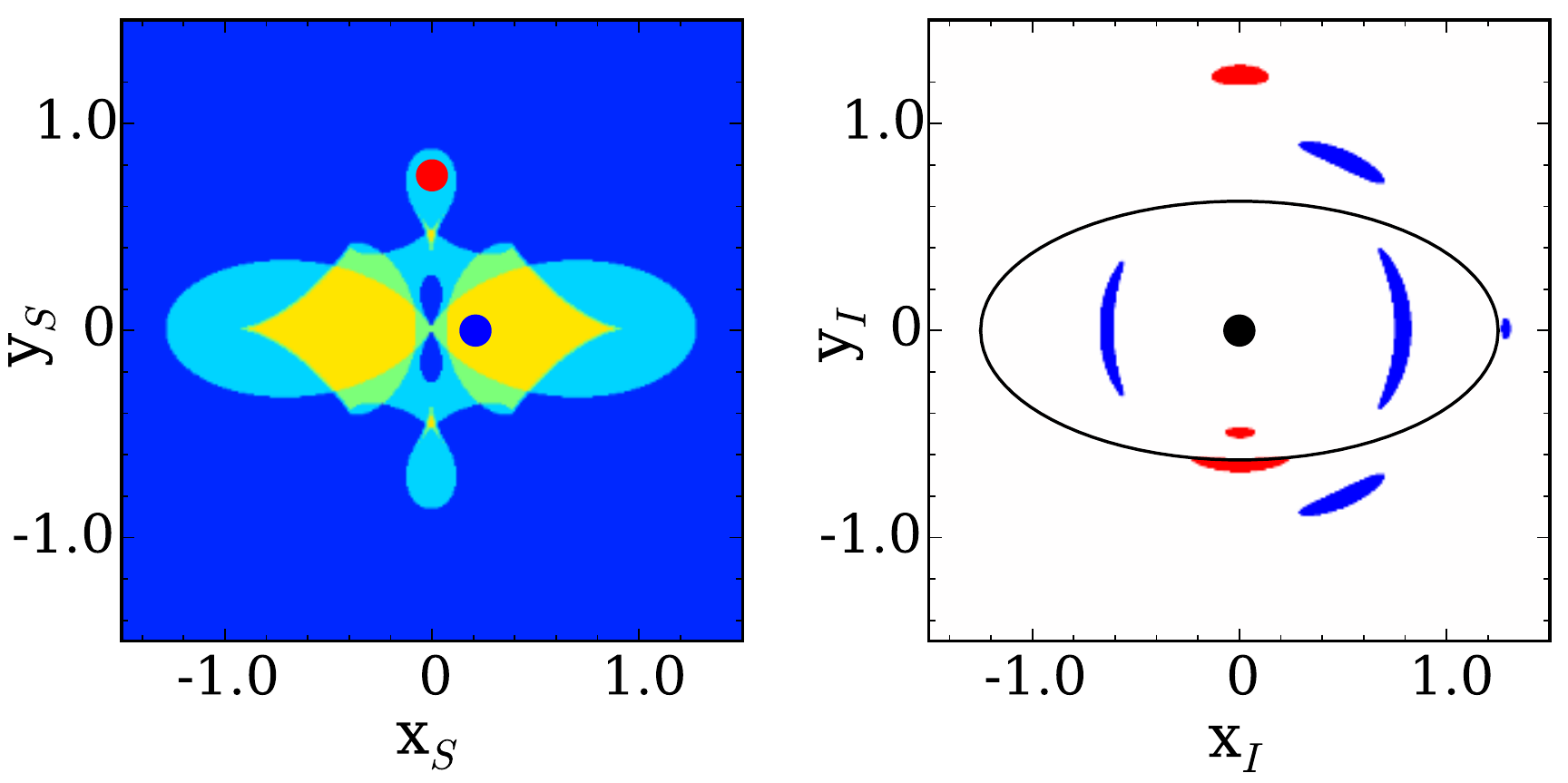}
\caption{
\label{fig:ring_mapping}
A demonstration of the lens mapping for two sample sources for a ring+point 
lens with the ring and the point having equal mass. The ring has a radius
$a = 1.25$ and is seen face-on for the top panels and inclined (with 
inclination angle $i=60^\circ$) for the bottom panels.
The left panels show image multiplicities in the source plane, with two 
sources plotted as the red and blue circles. The right panels show how these 
sources appear in the image plane, with the ring+point system drawn in black.
}
\end{figure*}

For the belt with radius centred at $a = 1.25$ (middle plot of 
Figure~\ref{fig:8panel_beltstar}), the magnification pattern shares some broad
similarities with that of the $a = 1.25$ ring case (bottom plot of
Figure~\ref{fig:8panel_ringstar}). The main noticeable differences are that at 
low inclination the belt+point case has a narrower region bounded by \pss\
and that in the face-on case, there is no image number change when crossing 
the \pss\ (for the same reason as in the above $a = 0.75$ case).
In particular, at edge on, the three cases from $a=0.75$ to
$a=1.25$ are similar to the triple point lensing system going from close toward 
intermediate separations. The overall features for the case with half mass 
inside and half mass outside the Einstein ring radius are in between the 
$a = 0.75$ and $a = 1.25$ cases.

\subsubsection{Examples of Images from the Ring+Point System}

In Figure~\ref{fig:ring_mapping}, we provide examples of the sorts of images 
formed for sources lensed by ring+point mass systems. The top plot shows the case 
of a face-on ring with $a = 1.25$. The left panel shows the image multiplicity 
map with the red and blue circles representing two sources in the source 
plane, and the corresponding images are shown in the image plane in the right 
panel. The red source is mapped to two images by the centre point mass, while 
the blue source inside the region bounded by the \pss\ is able to pass rays 
outside the ring (black circle) and take advantage of the ring's lensing 
potential to form three images, with the third appearing just outside the ring. 

The bottom plot considers a slightly more complicated case, with the ring 
inclined to $i = 60^{\circ}$. Even though the image multiplicity map in the
left panel looks complicated, the way in which each source is lensed is easy 
to understand. The red source is lensed by the centre point mass, but only one
of the two possible images is inside the ring and hence is physical. The
two images outside the ring come from rays deflected by both the point mass
and the ring.  The blue source can create two images inside the ring from
deflection solely by the centre point mass. On the $x_I$ axis, it can use 
additionally the ring's deflection to create an image just outside the ring.
Furthermore, it can create two additional images outside the ring by taking 
advantage of the $y_I$ component of the deflection. In total, five images 
emerge for this source.

\subsection{Smooth Density Distributions and the Origin of Pseudo-Caustics} 
\label{sec:smooth_distributions}

\begin{figure*}
\includegraphics[scale=.5]{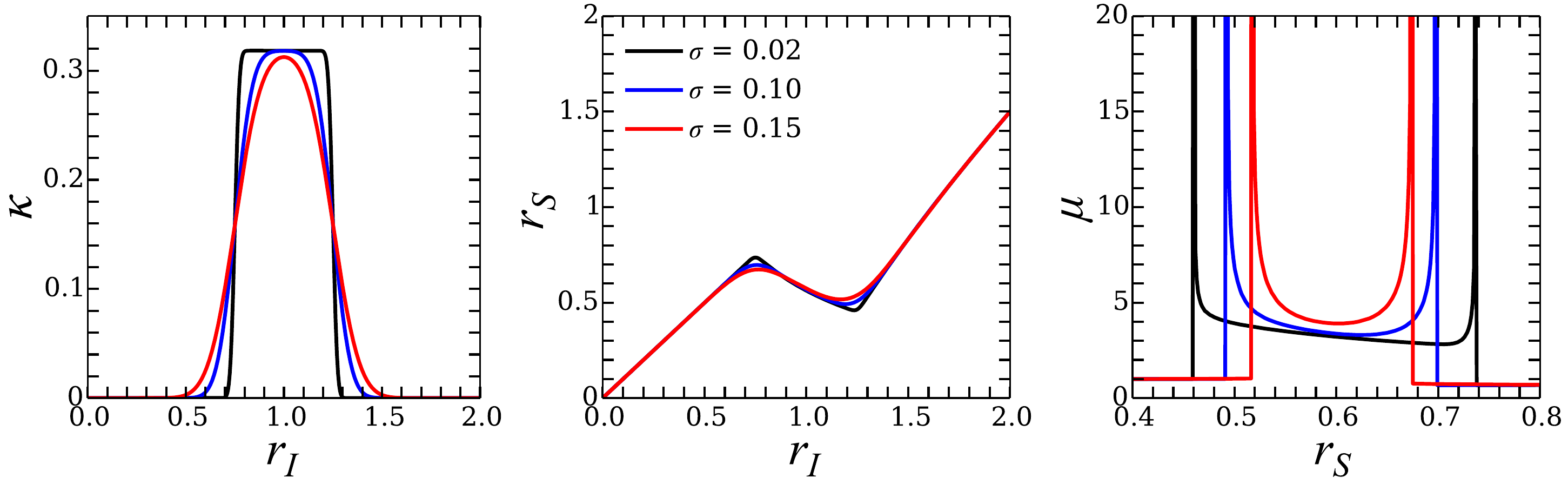}
\includegraphics[scale=.5]{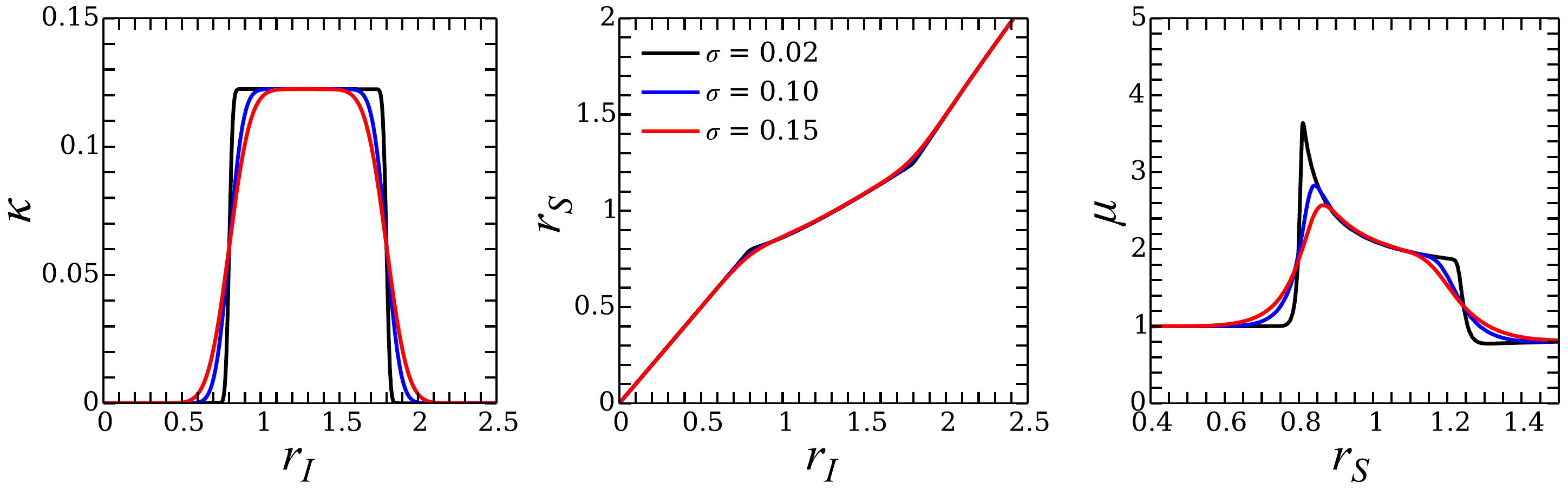}
\caption{\label{fig:faceon_smooth} The effects of smoothing out the mass distribution for face-on belts. From left to right, the columns show the surface density profile $\kappa(r_I)$ of the belt, the solution to the lensing equation ($r_S$--$r_I$ mapping), and the magnification $\mu(r_s)$ for three different values of the smoothing parameter $\sigma$. 
In the top rows, the belt has $a_c = 1$ and $\Delta a = 0.5$, so that $\Delta a < (\Delta a)_{\rm cr}$, meaning that the belt can form multiple images.
The \pss\ (associated with image number change) in the limit of 
$\sigma\rightarrow 0$ for such a case originates from disappearing formal 
caustics.
In the bottom row the belt has $a_c = 1.3$ and $\Delta a = 1$ so that $\Delta a > (\Delta a)_{\rm cr}$, and the belt forms a single image for all $r_S$. 
The \pss\ (associated with image size change) in the limit of 
$\sigma\rightarrow 0$ for such a case comes from sharpening the edge of
the bump in the magnification map. See the text for detail.
}
\end{figure*}

The belt/ring+point systems considered so far have all possessed discontinuous mass distributions. This 
allows for a simple analytic treatment of such systems, but in order to connect our results with
potential astrophysical applications (namely microlensing events caused by asteroid belt + star lenses), 
we need to consider a more realistic
smoothened mass distribution. { Working with a smooth mass distribution means 
that the conditions necessary for applying the relevant
theorems from singularity theory regarding image multiplicities and caustic morphologies are satisfied,
and as such we expect the \pss\ features to change. } As we will see,
by smoothing out the mass distribution, \pss\ either change into formal
caustics or become smeared into regions of finite but smooth magnification 
jumps, with the former corresponding to \pss\ associated with image number 
change and the latter with image size change. This provides a clear picture of
the origin of the \pss\ encountered in the systems considered in this paper 
with discontinuous mass distributions. 

To illustrate the effects of smoothing, we analyze a simple face-on belt-only model with a smoothed mass distribution of the form
\be 
\kappa(r_I) = \frac{1}{N}\left[\text{erf}\left(\frac{r_I-a_i}{\sigma}\right) - \text{erf}\left(\frac{r_I-a_o}{\sigma}\right)\right],
\ee
where $N$ normalizes the distribution. It represents a belt of characteristic 
inner radius $a_i$ and outer radius $a_o$, with the parameter $\sigma$ 
controlling the smoothness of the inner and outer edges. The belt systems studied
in previous sections correspond to the limit of $\sigma\rightarrow 0$.

The precise effect of the smoothing on the \pss\ depends on the size of the 
smoothing parameter $\sigma$ and the ratio $\Delta a / (\Delta a)_{\rm cr}$, 
where $(\Delta a)_{\rm cr} = 2/(a_o + a_i)$ [equation~(\ref{eqn:delta_a})]
is the critical belt width. As discussed earlier, sharp belts with $\sigma = 0$ and 
$\Delta a > (\Delta a)_{\rm cr}$ can generate \pss\ that result in 
changes in the sizes of images. If the width of the belt satisfies $\Delta a < 
(\Delta a)_{\rm cr}$, the belt can produce \pss\ that correspond to changes in
the image multiplicity. In these two regimes, the effects of smoothing are 
qualitatively different. This is illustrated in Figure~\ref{fig:faceon_smooth}. In each column
of the plot, from left to right the three panels show the surface density profile of the belt, the 
solution to the lensing equation $r_S(r_I)$, and the magnification as a function of $r_S$, 
respectively, for three values of $\sigma$ (0.02, 0.10, and 0.15).

In the top three panels, the belt has $\Delta a < (\Delta a)_{\rm cr}$. 
The smooth mass distribution of the belt leads to a smooth $r_S(r_I)$
mapping, and the \pss\ associated with a change in image multiplicity transform into
circular formal caustics with divergent $\mu$. As $\sigma$ increases, the 
two formal caustics move closer together in the source plane and the 
characteristic width of the caustic feature (defined as the extent in 
the source plane with magnification $\mu$ above some large threshold value) increases. 
Conversely, as we decrease $\sigma$, the extent of the caustic feature
shrinks. In the limit of $\sigma\rightarrow 0$ it shrinks to zero and the
caustic feature formally disappears. In this limit, the magnification shows a sharp finite
jump at the position of each former caustic, and the feature is 
identified as a \ps. So in the $\Delta a < (\Delta a)_{\rm cr}$ regime, 
the \pss\ are in fact associated with infinitesimally small formal caustics.

In the bottom three panels, the belt is wide, with 
$\Delta a > (\Delta a)_{\rm cr}$. When $\sigma \rightarrow 0$, the system possesses \pss\ 
associated with image size changes induced by sharp changes in $\partial r_S / \partial r_I$, 
which create finite jumps in $\mu$. Smoothing the mass distribution 
softens the jumps in $\mu$, and the \pss\ become smeared out. 
We can thus think about the origin of the \pss\ in the following way: for a smooth
mass distribution with $\sigma \ll \Delta a$, $\partial r_S / \partial r_I$
changes smoothly over small annular regions surrounding the belt boundaries. These changes translate to smooth magnification jumps
within narrow annular regions of the source plane. As $\sigma\rightarrow 0$, these jumps in $\mu$ become
sharp while remaining finite. The loci where the finite jumps occur are identified
as \pss.

\section{Summary and Discussion} 

In this paper, we have explored the gravitational lensing properties of a lens
made of a circular ring/belt, with and without a centre point mass. These 
systems exhibit complex features that vary significantly within the parameter space of the ring/belt, 
including open caustics and \pss. We detail a few examples 
that illustrate these features. 

The key property of a circular ring/belt lenses is that the rays passing 
interior to the projected ring/belt are not deflected by the ring/belt, while 
those passing exterior to the projected ring/belt are deflected. We can thus consider 
two types of solutions to the lensing equation -- one involves deflection from the entire 
ring/belt system, which we call the exterior solution, and one involves no deflection 
from the ring/belt, which we call the interior solution. The projected ring/belt thus acts like a filter 
to select which one of these solutions is applicable at different points in the image plane. That is, in the image plane, the ring/belt sets the boundary to 
determine which images are physical. 
Some solutions are forbidden, because they are obtained as interior solutions 
but correspond to rays passing exterior to the 
ring/belt, or vice versa. 
Because 
of this selection process, for the critical curves associated with the exterior 
solutions, only the part exterior to the belt/ring remains, and similarly for 
those associated with the interior solutions. This allows for broken critical curves, 
and hence open caustics.

In the image plane, the ring/belt is the region that connects the interior and 
exterior solutions. The connection can be continuous in the belt case and 
discontinuous in the thin ring case. In either case, the joint 
solution to the lens equation is generically non-differentiable at the boundaries of the belt/ring.
Since in general the boundaries of the belt/ring are not related to the
critical curves, the boundaries are not associated with the singularities
in the lensing mapping. The corresponding curves in the 
source plane (obtained by mapping the belt/ring boundaries into the source plane) 
are not associated with formal caustics (which have infinite 
magnification), but although there exists a finite discontinuity in magnification across
such curves. As such, they are dubbed ``\pss''. Such a magnification 
change can be
associated either with a change in the number of images formed by a source (e.g. for thin rings 
or narrow belts) or not (e.g. in the case of an extended belt; see the example
in Fig.~\ref{fig:face_on_belt}). The magnification change in the latter 
case is caused by a sharp change in the size of an image. These effects are present because of the 
discontinuous nature of the mass distribution of the ring/belt. When 
the density distribution is smoothened, the \pss\ become formal caustics if 
the belt is narrow, and disappear (or more precisely, the finite magnification 
jumps associated with the \pss\ become ``smeared out'') 
if the belt is wide. That is, 
the \pss\ associated with changes in image multiplicity are attributed to infinitely-thin 
formal caustics, and those associated with sharp changes in the size of an image are 
attributed to the sharpening of the edges of a smooth bump in the magnification map.

The images modified by the boundaries of ring/belt in the image plane and the 
magnification patterns modified by the \pss\ in the source plane form the main 
lensing properties of the lenses considered in this paper. We 
see a complex interplay between the ring/belt and the critical curves in the 
image plane, or between the \pss\ and the formal caustics in the source plane. 
Usually, analysing the (formal) caustics of a lensing system provides enough 
information to determine the image multiplicities, the general magnification 
patterns, and the light curve properties expected for a given source trajectory. 
However, once a ring/belt is introduced to the lensing system, in addition 
to the formal caustics, an analysis of the \pss\ is needed to obtain a full picture 
of the lensing properties of the system. A vast amount of information hidden 
at finite magnifications would be missed if a study concerning such systems
were to focus only on the formal 
caustics. Our work thus serves as a cautionary tale for investigations into other 
novel lensing geometries. 

The deflection and magnification of the lens mappings we have considered 
can be computed analytically, and the pseudo-caustics can 
be easily found by mapping the edges of the ring/belt to the source plane. However,
finding analytic solutions to the lensing equation is fairly tedious with 
our current approach, and we have been forced to restrict our analytic analysis to the simplest 
cases of face-on and edge-on systems (for a discussion of the latter, see Appendix~\ref{sec:appendix_edgeon}). 
Additionally, we lack an analytic expression for the image 
multiplicity at an arbitrary point in the source plane, which currently must be found 
numerically. A more general mathematical understanding of the connection between
\pss\ and image multiplicities would be of great utility to the theoretical study of gravitational lensing.

Although our analysis is mainly from a theoretical viewpoint and aims at 
understanding the main features in a lensing system with rings/belts, there are also 
potential applications of our work to astronomical situations. In a related 
paper \citep{Lake16}, 
we discuss the possibility of detecting extrasolar asteroid belts 
based on their microlensing signatures. For the lensing system of interest, 
the main difference lies in the mass ratio of the belt to the centre point 
(star) -- realistic mass ratios are around $q \sim 10^{-6}$ or smaller, as 
opposed to the $q = 1$ case 
considered here. However, it is shown that the \pss\ still have a significant effect 
on the magnification maps. This gives rise to interesting features in 
microlensing light curves that can potentially be used to detect extrasolar asteroid belts.

{ 
The other possible application is the lensing by ring galaxies. As a result
of head-on collisions between a disk galaxy and another galaxy 
\citep[e.g.][]{Lynds76,Appleton96}, a radially expanding density wave is 
excited in the disk galaxy, which triggers star formation. The fraction of 
stellar mass resultant from the star formation is at the sub-percent level
\citep[e.g.][]{Fogarty11}, and the total perturbed mass in the ring can be
even higher. If a background source (galaxy or quasar) happens to be lensed
by a ring galaxy, the observation of such a lensing system can be used to 
constrain the mass in the star-forming ring, which can be used to test the
density wave model of star formation and aid the understanding of ring
galaxies. 
}

\section*{ACKNOWLEDGEMENTS}
We thank the anonymous referees for constructive comments and suggestions that 
helped improve the paper. The support and resources from the centre for High
Performance Computing at the University of Utah are gratefully acknowledged.

\bibliographystyle{mnras}
\bibliography{ms}

\appendix

\section{Analysis of Edge-on Rings}

\label{sec:appendix_edgeon}
While a general analytic analysis for the inclined systems considered in the text is possible, the equations quickly become unwieldy. Thus, it is helpful to study the 
limiting cases of edge-on and face-on systems analytically in order to build an intuition for 
how the general cases behave. In this appendix, we present a more quantitative understanding 
for how edge-on rings behave as lenses, having treated the simpler face-on case in the main text (Section~\ref{sec:face_on_sys}).  
For simplicity, we only consider sources located along either the $x_S$ or $y_S$ axes. 

First, let us assume a source located along the $y_S$ axis, with $\bvr_S = (0,y_S)$ for $y_S > 0$. There are two types of images -- those located along the $y_I$ axis ($\bvr_I = (0,y_I)$), and those with a nonzero $x_I$ component. Starting from equation~(\ref{eqn:alpha_ring}), we can derive that the images along the $y_I$ axis are found by solving
\be \label{eqn:yaxis_soln} (y_I-y_S)^2(y_I^2+a^2) -1 =0.\ee
This equation generically has two real solutions for $y_I$. For small $y_S$ (more precisely, for $y_S < 1/a$), there are both positive and negative 
solutions for $y_I$, corresponding to rays passing ``over'' or ``under'' the ring. As $y_S$ increases past the critical 
value of $1/a$, the ``under the ring'' solution 
occurs at $y_I>0$, 
and hence becomes unphysical. 

The other images with $x_I \neq 0 $ are found by solving the equation
\be \label{eq:edgeonxneq0soln} 1-y_S/y_I = (1-y_I/y_S)^2(a^2+y_S^2)^2 \ee
for $y_I$, which will generically have three real solutions. After obtaining $y_I$, we obtain two symmetric solutions for $x_I$ by 
\be x_I = \pm \sqrt{y^2(y_S/y_I-1) - a^2(y_I/y_S-1)}. \ee
Equation~(\ref{eq:edgeonxneq0soln}) is cubic, and so we can generically get three real solutions. However, only negative $y_I$ solutions to equation~(\ref{eq:edgeonxneq0soln}) are physical. To see this, after a little manipulating of the lens equation, we obtain $\lambda = -a^2 y_I / y_S$, and since
$\lambda$ is always positive, any images with $x_I \neq 0$ must have a $y_I$ coordinate opposite in sign to $y_S$, which means that we always get two such images, both with the same negative $y_I$-coordinate and with opposite $x_I$ coordinates. 

This analysis allows us to understand why even very highly misaligned sources located along the $y_S$ axis can generate 3-images in the edge-on case (see the rightmost column of Figure~\ref{fig:8panel_ringonly}). One of the three images is located at $x_I = 0$ and $y_I > 0$, while the other two are at located at $y_I < 0$ and related to each other by reflection about the $y_I$ axis. At smaller $y_S$, the source creates four images, with the transition between the 3-image region and the 4-image region occurring at a critical source position of $y_{S,{\rm cr}} = a^{-1}$, below which equation~(\ref{eqn:yaxis_soln}) has both positive and negative real solutions. 

We now turn to diagnosing the situation for images along the $x_S$ axis. The general results are similar to those for images located along the $y_S$ axis. By beginning from the lens equation, we find images along that are located along the $x_I$ axis by solving 
\be (x_I-x_S)^2(x_I^2-a^2) = 1. \ee
Note that this means that we only can form images along the $x_I$ axis if $x_I>a$. For images with nonzero $y_I$, we find $x_I$ by solving
\be 1 - x_S/x_I = (x_I/x_S-1)^2(a^2-x_S^2)^2. \ee
Once we obtain $x_I$, we obtain two mirror symmetric solutions for $y_I$ given by 
\be y_I = \pm \sqrt{ x_I^2(x_S/x_I -1) + a^2(x_I /x_S - 1)}. \ee

\section{Analytic Calculation of the Magnification for the Ring/Belt+Point System} \label{sec:analytic_mu_appendix}

The magnification for a ring/belt+point system at a given point in the source plane can be obtained analytically, which is very useful for 
performing numerical calculations and identifying the caustic set of a given lensing geometry. The magnification at a point $\bvec{r}_I$ in the image plane is
\be
\mu (\bvec{r}_I) = \frac{1}{\textrm{det}(A)},
\ee
where the Jacobian of the lens mapping is given by 
\be
A_{ij} = \delta_{ij} - \frac{\partial \alpha_i}{\partial x_j}.
\ee
For the ring+point mass system, the Jacobian matrix is
\be \label{eq:mag}
A= \left( \begin{array}{cc}
1 - g_{xx} & g_{xy} \\
g_{yx} & 1 - g_{yy} \end{array} \right) ,
\ee
where 
\be\label{eq:terms_in_Jacobian}
\begin{aligned}
g_{xx} & =  \frac{1}{1+q}\frac{|\bvec{r}_I|^2 - 2x_I^2}{|\bvec{r}_I|^4} + \frac{q}{1+q}\frac{\partial \alpha_{r,x}}{\partial x_I}\Theta\left(\frac{x_I^2}{a^2}+\frac{y_I^2}{b^2} - 1 \right), \\ 
g_{xy} & = g_{yx} = -\frac{1}{1+q}\frac{2x_Iy_I}{|\bvec{r}_I|^4} + \frac{q}{1+q}\frac{\partial \alpha_{r,x}}{\partial y_I}\Theta\left(\frac{x_I^2}{a^2}+\frac{y_I^2}{b^2} - 1 \right), \\
g_{yy}  & =  \frac{1}{1+q}\frac{|\bvec{r}_I|^2 - 2y_I^2}{|\bvec{r}_I|^4} + \frac{q}{1+q}\frac{\partial \alpha_{r,y}}{\partial y_I}\Theta\left(\frac{x_I^2}{a^2}+\frac{y_I^2}{b^2} - 1 \right). 
\end{aligned}
\ee
The first terms in each equation are from the point mass, with the second terms coming from the ring's contribution. Here, $\Theta$ is the Heaviside step function and $\alpha_{r,x}$ ($\alpha_{r,y}$) denotes the $x$-component ($y$-component) of the deflection angle due to the ring. The partial derivatives of the deflection angle in equation~(\ref{eq:terms_in_Jacobian}) can be computed analytically, yielding
\be\label{eqn:deriv_ring}
\begin{aligned}
\frac{\partial \alpha_{r,x}}{\partial x_I} & = \frac{p'^2}{a'^5b'}\left(a'^2-x_I\frac{\partial \lambda}{\partial x_I}\right) 
	+ \frac{2p'^4x_I}{a'^3b'} \left[ \left(\frac{x_I^2}{a'^6}+\frac{y_I^2}{b'^6}\right) \frac{\partial \lambda}{\partial x_I} -\frac{x_I}{a'^4} \right] \\ & \qquad-\frac{p'^2x_I}{a'^4b'^2}\frac{\partial \lambda}{\partial x_I} \left( \frac{b'}{2a'} + \frac{a'}{2b'} \right), \\
\frac{\partial \alpha_{r,y}}{\partial y_I} & = \frac{p'^2}{a'b'^5}\left(b'^2-y_I\frac{\partial \lambda}{\partial y_I}\right)
	+ \frac{2p'^4y_I}{a'b'^3}\left[ \left(\frac{x_I^2}{a'^6}+\frac{y_I^2}{b'^6}\right) \frac{\partial \lambda}{\partial y_I} -\frac{y_I}{b'^4} \right]  \\ & \qquad - \frac{p'^2y_I}{a'^2b'^4}\frac{\partial \lambda}{\partial y_I} \left( \frac{b'}{2a'} + \frac{a'}{2b'} \right), \\ 
\frac{\partial \alpha_{r,x}}{\partial y_I} & = \frac{\partial \alpha_y}{\partial x_I} = -\frac{p'^2x_I}{a'^5b'}\frac{\partial \lambda}{\partial y_I} 
	+ \frac{2p'^4x_I}{a'^3b'}\left[ \left(\frac{x_I^2}{a'^6}+\frac{y_I^2}{b'^6}\right) \frac{\partial \lambda}{\partial y_I}  -\frac{y_I}{b'^4} \right]   \\ & \qquad  - \frac{p'^2x_I}{a'^4b'^2}\frac{\partial \lambda}{\partial y_I} \left( \frac{b'}{2a'} + \frac{a'}{2b'} \right),
\end{aligned}
\ee
where the primed variables retain their meanings from section \ref{sec:inclined_systems} [see equation~(\ref{eq:confocal_ellipse})]. The partial derivatives of $\lambda$ are given by 
\be
\frac{\partial \lambda}{\partial x_I} = \frac{2p'^2x_I}{a'^2} \quad {\rm and} \quad \frac{\partial \lambda}{\partial y_I} = \frac{2p'^2y_I}{b'^2}.
\ee

An analytic expression also exists for thick belts. Letting $a_i$ and $a_o$ be the inner and outer radii (respectively) of the belt, we compute the magnification as in equation~(\ref{eq:terms_in_Jacobian}), but this time with the factors of $\Theta\left({x_I^2/a^2}+{y_I^2/b^2} - 1 \right)$ replaced by $\Theta\left({x_I^2/a_i^2}+{y_I^2/b_i^2} - 1 \right)$, and with the partial derivatives replaced with 
\be\begin{aligned}
\label{eqn:deriv_belt}
\frac{\partial \alpha_{b,x}}{\partial x_I} & = 2a_ob_o\left(\frac{1}{a_o'(a_o'+b_o')} - \frac{\partial \lambda_o}{\partial x_I}\frac{x_I}{2a_o'^3b_o'}\right) \\ & \qquad- 
	 2a_ib_i\left(\frac{1}{a_i'(a_i'+b_i')} - \frac{\partial \lambda_i}{\partial x_I}\frac{x_I}{2a_i'^3b_i'}\right), \\
\frac{\partial \alpha_{b,y}}{\partial y_I} & = 2a_ob_o\left(\frac{1}{b_o'(a_o'+b_o')} - \frac{\partial \lambda_o}{\partial y_I}\frac{y_I}{2a_o'b_o'^3}\right)\\ & \qquad - 
	 2a_ib_i\left(\frac{1}{b_i'(a_i'+b_i')} - \frac{\partial \lambda_i}{\partial y_I}\frac{y_I}{2a_i'b_i'^3}\right), \\
\frac{\partial \alpha_{b,x}}{\partial y_I} & = \frac{\partial \alpha_y}{\partial x_I} = 2a_ob_o\left( \frac{\partial \lambda_o}{\partial y_I}\frac{-x_I}{2a_o'^3b_o'}\right) 
	 - 2a_ib_i\left( \frac{\partial \lambda_i}{\partial y_I}\frac{-x_I}{2a_i'^3b_i'}\right).
\end{aligned} \ee
The primed semi-axes are defined through
\be
a_i'^2 = a_i^2 + \lambda_i, \quad b_i'^2 = b_i^2 + \lambda_i, \quad a_o'^2 = a_o^2 + \lambda_o, \quad b_o'^2 = b_o^2 + \lambda_o,
\ee
and 
\be
p_i' = \left( x_I^2/a_i'^4 + y_I^2/b_i'^4 \right)^{-1/2}, \quad p_o' = \left( x_I^2/a_o'^4 + y_I^2/b_o'^4 \right)^{-1/2}\ee
are the perpendicular distances from the center of the inner and outer confocal ellipses to the tangent lines at $\bvr_I$, respectively. 

As in the thin ring case, $\lambda_i$ ($\lambda_o$) is found by solving 
equation~(\ref{eq:confocal_ellipse}) and using $a = a_i$ and $b=b_i$ ($a=a_o$ and $b=b_o$). The partial derivatives of $\lambda_i$ are 
\be
\frac{\partial \lambda_i}{\partial x_I} = \frac{2p_i'^2x_I}{a_i'^2} \quad {\rm and}
	\quad \frac{\partial \lambda_i}{\partial y_I} = \frac{2p_i'^2y_I}{b_i'^2},
\ee
while those of $\lambda_o$ are 
\be
 \frac{\partial \lambda_o}{\partial x_I} = \frac{2p_o'^2x_I}{a_o^2}\Theta\left(\frac{x_I^2}{a_o^2}+\frac{y_I^2}{b_o^2} - 1 \right) \quad {\rm and}
	\quad \frac{\partial \lambda_o}{\partial y_I} = \frac{2p_o'^2y_I}{b_o'^2}\Theta\left(\frac{x_I^2}{a_o^2}+\frac{y_I^2}{b_o^2} - 1 \right).
\ee

We note that for a ring or disc with uniform density 
\citet{Blandford91} derive the gradient of the deflection angle with an 
alternative method.

\section{Explicit Expression for the Pseudo-Caustics}
\label{sec:pss_expression}

We now present an explicit expression for the \pss\ of a belt+point mass system at arbitrary inclination by mapping the inner and outer edge of the belt to 
the source plane. There are two sets of pseudo-caustics, which we will label as $\mathbfcal{PS}_1$ and $\mathbfcal{PS}_2$. They are parametrized by an angle $\phi \in [0,2\pi)$ and given by 
\be \ba \mathbfcal{PS}_1(\phi)  &= (a_i\cos\phi \bvec{\hat x_S} + b_i\sin\phi \bvec{\hat y_S})\left(1 - \frac{1}{1+q}\frac{1}{a_i^2\cos^2\phi + b_i^2\sin^2\phi}\right) \ea \ee
and
\be \ba & \mathbfcal{PS}_2(\phi)  = a_o\cos\phi \bvec{\hat x_S} \bigg[1-\frac{1}{(1+q)(a_o^2\cos^2\phi + b_o^2\sin^2\phi)}   \\ & - \frac{q}{(1+q)(a_ob_o - a_ib_i)}\bigg(\frac{2b_o}{a_o+b_o} - \frac{2a_ib_i}{(a_i'(\phi)+b_i'(\phi))a_i'(\phi)}\bigg)\bigg]   
\\ & + b_o\sin\phi \bvec{\hat y_S} \bigg[1-\frac{1}{(1+q)(a_o^2\cos^2\phi + b_o^2\sin^2\phi)} \\&  - \frac{q}{(1+q)(a_ob_o - a_ib_i)}\bigg(\frac{2a_o}{a_o+b_o} - \frac{2a_ib_i}{(a_i'(\phi)+b_i'(\phi))b_i'(\phi)}\bigg)\bigg], \ea \ee
where
\be a_i'(\phi) = \sqrt{a_i^2 + \lambda_i(\phi)}\quad {\rm and}\quad b_i'(\phi) = \sqrt{b_i^2 + \lambda_i(\phi)} \ee
with 
\be \ba \lambda_i(\phi) & = \frac{a_o^2\cos^2\phi + b_o^2\sin^2\phi - a_i^2-b_i^2}{2} \\ & +\bigg[ \frac{1}{4}\left(a_o^2\cos^2\phi + b_o^2\sin^2\phi - a_i^2-b_i^2\right)^2  \\ &- a_i^2b_i^2 + b_i^2a_o^2\cos^2\phi + a_i^2b_o^2\sin^2\phi\bigg]^{1/2}.\ea \ee

\end{document}